%

\input harvmac
\input epsf

\newcount\figno
\figno=0

\def\fig#1#2#3{
\par\begingroup\parindent=0pt\leftskip=1cm\rightskip=1cm\parindent=0pt
\baselineskip=11pt \global\advance\figno by 1 \midinsert
\epsfxsize=#3 \centerline{\epsfbox{#2}} \vskip 12pt {\bf Fig.\
\the\figno: } #1\par
\endinsert\endgroup\par
}
\def\figlabel#1{\xdef#1{\the\figno}}
\def\encadremath#1{\vbox{\hrule\hbox{\vrule\kern8pt\vbox{\kern8pt
\hbox{$\displaystyle #1$}\kern8pt} \kern8pt\vrule}\hrule}}

\def\fullfig#1#2#3{
\par\begingroup\parindent=0pt\leftskip=1cm\rightskip=1cm\parindent=0pt
\baselineskip=11pt \global\advance\figno by 1 \pageinsert
\epsfxsize=#3 \centerline{\epsfbox{#2}} \vskip 12pt {\bf Fig.\
\the\figno: } #1\par
\endinsert\endgroup\par
}
\def\figlabel#1{\xdef#1{\the\figno}}
\def\encadremath#1{\vbox{\hrule\hbox{\vrule\kern8pt\vbox{\kern8pt
\hbox{$\displaystyle #1$}\kern8pt} \kern8pt\vrule}\hrule}}

\def\lesssim{\mathrel{\mathpalette\fun <}}
\def\gtrsim{\mathrel{\mathpalette\fun >}}
\def\fun#1#2{\lower3.6pt\vbox{\baselineskip0pt\lineskip.9pt
  \ialign{$\mathsurround=0pt#1\hfil##\hfil$\crcr#2\crcr\sim\crcr}}}
\relax

\lref\nf{S. Dimopoulos, S. Kachru, J. McGreevy, and J.G. Wacker,
 ``N-flation,'' hep-th/0507205.}

\lref\kklt{S. Kachru, R. Kallosh, A. Linde, and S. Trivedi, ``de
Sitter Vacua in String Theory,'' Phys. Rev. D {\bf 68} (2003)
046005, hep-th/0301240.}

\lref\GKP{S. Giddings, S. Kachru, and J. Polchinski, ``Hierarchies
from Fluxes in String Compactifications,'' Phys. Rev. D {\bf 66}
(2002) 106006, hep-th/0105097.}

\lref\GVW{S.~Gukov, C.~Vafa, and E.~Witten, ``CFT's From
Calabi-Yau Four-folds,'' Nucl.\ Phys.\ B {\bf 584} (2000) 69,
hep-th/9906070.}

\lref\Witten{E.~Witten, ``Non-Perturbative Superpotentials In
String Theory,'' Nucl.\ Phys.\ B {\bf 474}, 343 (1996),
hep-th/9604030.
}

\lref\assist{
  A.~R.~Liddle, A.~Mazumdar and F.~E.~Schunck,
  ``Assisted inflation,''
  Phys.\ Rev.\ D {\bf 58}, 061301 (1998), astro-ph/9804177.
}

\lref\littleinflaton{
  D.~E.~Kaplan and N.~J.~Weiner,
  ``Little inflatons and gauge inflation,''
  JCAP {\bf 0402}, 005 (2004), hep-ph/0302014.
}

\lref\chaotic{
   A.~D.~Linde,
   ``Chaotic Inflation,''
   Phys.\ Lett.\ B {\bf 129}, 177 (1983).
}

\lref\douglas{
 M.~R.~Douglas,
  ``The statistics of string / M theory vacua,''
  JHEP {\bf 0305}, 046 (2003), hep-th/0303194\semi
 S.~Ashok and M.~R.~Douglas,
  ``Counting flux vacua,''
  JHEP {\bf 0401}, 060 (2004), hep-th/0307049\semi
  F.~Denef and M.~R.~Douglas,
  ``Distributions of flux vacua,''
  JHEP {\bf 0405}, 072 (2004), hep-th/0404116\semi
  F.~Denef and M.~R.~Douglas,
  ``Distributions of nonsupersymmetric flux vacua,''
  JHEP {\bf 0503}, 061 (2005), hep-th/0411183.
}

\lref\dine{
  M.~Dine and N.~Seiberg,
  ``Nonrenormalization Theorems In Superstring Theory,''
  Phys.\ Rev.\ Lett.\  {\bf 57}, 2625 (1986).
}

\lref\wishart{J. Wishart, ``The Generalised Product Moment
Distribution in Samples from a Normal Multivariate Population,''
Biometrika A {\bf{20}}, 32 (1928).}

\lref\corr{S. H. Simon and A.L. Moustakas, ``Eigenvalue Density of
Correlated Complex Random Wishart Matrices,'' Phys.\ Rev.\ E {\bf
69}, 065101 (2004), math-ph/0401038.}

\lref\BanksSX{
  T.~Banks, M.~Dine, P.~J.~Fox and E.~Gorbatov,
``On the possibility of large axion decay constants,''
  JCAP {\bf 0306}, 001 (2003), hep-th/0303252.
}

\lref\richardtoappear{
  A.~Aazami and R.~Easther, ``Cosmology From Random Multifield Potentials,''
  hep-th/0512050.
}

\lref\wittalk{E.~Witten, talk at Strings 2005\semi P.~Svr\v{c}ek
and E.~Witten, to appear.}

\lref\wignrev{E. Wigner, ``Random Matrices in Physics,'' SIAM
Review {\bf{9}}, 1 (1967).}

\lref\Kanti{
  P.~Kanti and K.~A.~Olive,
  ``Assisted chaotic inflation in higher dimensional theories,''
  Phys.\ Lett.\ B {\bf 464}, 192 (1999), hep-ph/9906331.
}

\lref\allassist{
  P.~Kanti and K.~A.~Olive,
  ``On the realization of assisted inflation,''
  Phys.\ Rev.\ D {\bf 60}, 043502 (1999), hep-ph/9903524\semi
  K.~A.~Malik and D.~Wands,
  ``Dynamics of assisted inflation,''
  Phys.\ Rev.\ D {\bf 59}, 123501 (1999), astro-ph/9812204.
}

\lref\kaloperliddle{
  N.~Kaloper and A.~R.~Liddle,
  ``Dynamics and perturbations in assisted chaotic inflation,''
  Phys.\ Rev.\ D {\bf 61}, 123513 (2000), hep-ph/9910499.
}


\lref\nottwoone{
  G.~Curio, A.~Krause and D.~Lust,
``Moduli stabilization in the heterotic / IIB discretuum,''
hep-th/0502168.
}

\lref\bp{R. Bousso and J. Polchinski, ``Quantization of Four-Form
Fluxes and Dynamical Neutralization of the Cosmological
Constant,'' JHEP {\bf 0006} (2000) 006, hep-th/0004134.}

\lref\twoa{
  O.~DeWolfe, A.~Giryavets, S.~Kachru and W.~Taylor,
  ``Type IIA moduli stabilization,''
  JHEP {\bf 0507}, 066 (2005), hep-th/0505160.
}

\lref\bart{ N.~Bartolo, S.~Matarrese and A.~Riotto,
  ``Adiabatic and isocurvature perturbations from inflation: Power spectra  and
  consistency relations,''
  Phys.\ Rev.\ D {\bf 64}, 123504 (2001), astro-ph/0107502.
}

 \lref\mathmat{D.~S.~Bernstein, {\it{Matrix Mathematics}},
Princeton University Press, Princeton, 2005.}

\lref\edel{ I. Dumitriu and A. Edelman, ``Global spectrum
fluctuations for the $\beta$-Hermite and $\beta$-Laguerre
ensembles via matrix models,'' math-ph/0510043.}

\lref\BDaxion{T.~Banks and M.~Dine, ``The cosmology of string
theoretic axions,'' Nucl.\ Phys.\ B {\bf 505} (1997) 445,
hep-th/9608197.}

\lref\fluxcpt{Flux compactification references.}

\lref\bhko{ M.~Berg, M.~Haack and B.~K\"ors, ``Loop Corrections to
Volume Moduli and Inflation in String Theory,'' Phys.\ Rev.\ D
{\bf 71}, 026005 (2005) hep-th/0404087.
}

\lref\BKQ{ C.~P.~Burgess, R.~Kallosh and F.~Quevedo, ``de Sitter
string vacua from supersymmetric D-terms,'' JHEP {\bf 0310}, 056
(2003), hep-th/0309187.
}

\lref\KaloperAJ{
  N.~Kaloper and L.~Sorbo,
  ``Of pNGB QuiNtessence,'' astro-ph/0511543.
}

\lref\mehta{ M.L. Mehta, {\it{Random Matrices}}, Academic Press,
Boston, 1990.}

\lref\mp{ V.A. Mar\v{c}enko and L.A. Pastur, ``Distributions of
Eigenvalues for Some Sets of Random Matrices,'' Math. USSR.
Sbornik {\bf 1}, 457 (1967). }

\lref\wign{E. Wigner, ``On the Distribution of the Roots of
Certain Symmetric Matrices,'' Ann. of Math. {\bf 67}, 325 (1958).}

\lref\hungar{F. Oravecz and D. Petz, ``On the eigenvalue
distribution of some symmetric random matrices,'' Acta. Sci. Math.
Szeged. {\bf 63}, 383 (1997).}

\lref\comm{ A.M. Tulino and S. Verd\'{u}, {\it{Random Matrix
Theory and Wireless Communications}}, Now Publishers, Hanover,
2004.}

\lref\bai{ Z.D. Bai, ``Methodologies in Spectral Analysis of Large
Dimensional Random Matrices, A Review,'' Statistica Sinica
{\bf{9}}, 611 (1999).}

\lref\Dodelson{S.~Dodelson, W.~H.~Kinney and E.~W.~Kolb, ``Cosmic
microwave background measurements can discriminate among inflation
models,'' Phys.\ Rev.\ D {\bf 56}, 3207 (1997), astro-ph/9702166.}

\lref\shifty{
  J.~P.~Hsu, R.~Kallosh and S.~Prokushkin,
  ``On brane inflation with volume stabilization,''
  JCAP {\bf 0312}, 009 (2003), hep-th/0311077\semi
H.~Firouzjahi and S.~H.~H.~Tye,
  ``Closer towards inflation in string theory,''
  Phys.\ Lett.\ B {\bf 584}, 147 (2004), hep-th/0312020\semi
  J.~P.~Hsu and R.~Kallosh,
  ``Volume stabilization and the origin of the inflaton shift symmetry in
  string theory,''
  JHEP {\bf 0404}, 042 (2004), hep-th/0402047\semi
  S.~E.~Shandera,
  ``Slow roll in brane inflation,''
  JCAP {\bf 0504}, 011 (2005), hep-th/0412077.
}

\lref\noshift{
M.~Berg, M.~Haack and B.~K\"ors, ``Loop Corrections to Volume
Moduli and Inflation in String Theory,'' Phys.\ Rev.\ D {\bf 71},
026005 (2005), hep-th/0404087\semi
M.~Berg, M.~Haack and B.~K\"ors, ``On the Moduli Dependence of
Nonperturbative Superpotentials in Brane Inflation,''
hep-th/0409282\semi
L.~McAllister, ``An inflaton mass problem in string inflation from
threshold corrections to volume stabilization,'' hep-th/0502001.
}

\lref\Kawasaki{
M.~Kawasaki, M.~Yamaguchi and T.~Yanagida,
   ``Natural chaotic inflation in supergravity,''
   Phys.\ Rev.\ Lett.\  {\bf 85}, 3572 (2000),
   hep-ph/0004243.}

\lref\Linde{ A.~Linde, ``Inflation and string cosmology,'' eConf
{\bf C040802}, L024 (2004), hep-th/0503195.}

\lref\Lyth{
   D.~H.~Lyth,
   ``What would we learn by detecting a gravitational wave signal in
the  cosmic
   microwave background anisotropy?,''
   Phys.\ Rev.\ Lett.\  {\bf 78}, 1861 (1997), hep-ph/9606387.}

\lref\Sasaki{
  M.~Sasaki and E.~D.~Stewart,
  ``A General Analytic Formula for the Spectral Index of the Density
  Perturbations produced during Inflation,''
  Prog.\ Theor.\ Phys.\  {\bf 95}, 71 (1996), astro-ph/9507001.
}

\lref\technicaladvance{
K.~Dasgupta, G.~Rajesh and S.~Sethi, ``M theory, orientifolds and
G-flux,'' JHEP {\bf 9908}, 023 (1999), hep-th/9908088\semi
E.~Silverstein, ``(A)dS backgrounds from asymmetric
orientifolds,'' hep-th/0106209\semi
A.~Maloney, E.~Silverstein and A.~Strominger, ``De Sitter space in
noncritical string theory,'' hep-th/0205316\semi
B.~S.~Acharya, ``A moduli fixing mechanism in M theory,''
hep-th/0212294\semi
  F.~Denef, M.~R.~Douglas and B.~Florea,
  ``Building a better racetrack,''
  JHEP {\bf 0406}, 034 (2004), hep-th/0404257\semi
  V.~Balasubramanian, P.~Berglund, J.~P.~Conlon and F.~Quevedo,
  ``Systematics of moduli stabilisation in Calabi-Yau flux compactifications,''
  JHEP {\bf 0503}, 007 (2005), hep-th/0502058\semi
  F.~Denef, M.~R.~Douglas, B.~Florea, A.~Grassi and S.~Kachru,
  ``Fixing all moduli in a simple F-theory compactification,''
  hep-th/0503124\semi
  P.~S.~Aspinwall and R.~Kallosh,
  ``Fixing all moduli for M-theory on K3 x K3,''
  hep-th/0506014.
}

\lref\stringinflation{P.~Binetruy and G.~R.~Dvali, ``D-term
inflation,'' Phys.\ Lett.\ B {\bf 388} (1996) 241,
hep-ph/9606342\semi G.~R.~Dvali and S.~H.~H.~Tye, ``Brane
inflation,'' Phys.\ Lett.\ B {\bf 450}, 72 (1999),
hep-ph/9812483\semi
S. Alexander, ``Inflation from D - Anti-D-Brane Annihilation,''
Phys. Rev. D {\bf 65} (2002) 023507, hep-th/0105032\semi G. Dvali,
Q. Shafi and S. Solganik, ``D-brane Inflation,''
hep-th/0105203\semi C.P. Burgess, M. Majumdar, D. Nolte, F.
Quevedo, G. Rajesh and R.J. Zhang, ``The Inflationary
Brane-Antibrane Universe,'' JHEP {\bf 07} (2001) 047,
hep-th/0105204\semi J. Garcia-Bellido, R. Rabadan and F. Zamora,
``Inflationary Scenarios from Branes at Angles,'' JHEP {\bf 0201}
(2002) 036, hep-th/0112147\semi K. Dasgupta, C. Herdeiro, S.
Hirano and R. Kallosh, ``D3/D7 Inflationary Model and M-theory,''
Phys. Rev. D {\bf 65} (2002) 126002, hep-th/0203019\semi
N.~Jones, H.~Stoica and S.~H.~H.~Tye, ``Brane interaction as the
origin of inflation,'' JHEP {\bf 0207}, 051 (2002),
hep-th/0203163\semi
S.~Kachru, R.~Kallosh, A.~Linde, J.~Maldacena, L.~McAllister and
S.~P.~Tri\-vedi, ``Towards inflation in string theory,'' JCAP {\bf
0310}, 013 (2003), hep-th/0308055 \semi
 E.~Silverstein and D.~Tong,
  ``Scalar speed limits and cosmology: Acceleration from D-cceleration,''
  Phys.\ Rev.\ D {\bf 70}, 103505 (2004), hep-th/0310221\semi
  O.~DeWolfe, S.~Kachru and H.~L.~Verlinde, ``The giant inflaton,''
  JHEP {\bf 0405}, 017 (2004), hep-th/0403123\semi
N.~Iizuka and S.~P.~Trivedi, ``An inflationary model in string
theory,''
  Phys.\ Rev.\ D {\bf 70}, 043519 (2004), hep-th/0403203\semi
  M.~Alishahiha, E.~Silverstein and D.~Tong,
  ``DBI in the sky,''
  Phys.\ Rev.\ D {\bf 70}, 123505 (2004), hep-th/0404084\semi
J.~J.~Blanco-Pillado {\it et al.}, ``Racetrack inflation,'' JHEP
{\bf 0411}, 063 (2004), hep-th/0406230\semi

 K.~Becker, M.~Becker and A.~Krause,
  ``M-theory inflation from multi M5-brane dynamics,''
  Nucl.\ Phys.\ B {\bf 715}, 349 (2005), hep-th/0501130\semi
  A.~Westphal,
  ``Eternal inflation with alpha' corrections,''
  JCAP {\bf 0511}, 003 (2005), hep-th/0507079\semi
  J.~P.~Conlon and F.~Quevedo,
  ``Kaehler moduli inflation,'' hep-th/0509012.
}

\lref\Jokinen{
  A.~Jokinen and A.~Mazumdar,
  ``Inflation in large N limit of supersymmetric gauge theories,''
  Phys.\ Lett.\ B {\bf 597}, 222 (2004), hep-th/0406074.
}

\lref\Alabidi{
  L.~Alabidi and D.~H.~Lyth, ``Inflation models and observation,''
  astro-ph/0510441.
}

\lref\evencorrelated{
  J.~H.~Schenker and H.~Schulz-Baldes, ``Semicircle law and
  freeness for random matrices with symmetries or correlations,''
  MRL {\bf 12}, 531 (2005), math-ph/0505003.
}

\lref\andreibook{ Andrei's book} \lref\liddlelyth{A.R. Liddle and
D.H. Lyth, {\it{Cosmological Inflation and Large-Scale
Structure}}, Cambridge University Press, Cambridge, 2000.}

\lref\bartjanphd{ B.J.W. van Tent, ``Cosmological Inflation with
Multiple Fields and the Theory of Density Fluctuations,'' Utrecht
University Ph.D. dissertation. }

\lref\banks{
  T.~Banks, M.~Dine and M.~Graesser,
  ``Supersymmetry, axions and cosmology,''
  Phys.\ Rev.\ D {\bf 68}, 075011 (2003), hep-ph/0210256.
}

\lref\uday{
  J.~Distler and U.~Varadarajan, ``Random polynomials and the friendly landscape,'' hep-th/0507090.
}

\lref\PeirisFF{ H.~V.~Peiris {\it et al.}, ``First year Wilkinson
Microwave Anisotropy Probe (WMAP) observations: Implications for
inflation,'' Astrophys.\ J.\ Suppl.\  {\bf 148}, 213 (2003),
astro-ph/0302225.
}
 
\lref\SpergelCB{ D.~N.~Spergel {\it et al.}  [WMAP Collaboration],
``First Year Wilkinson Microwave Anisotropy Probe (WMAP)
Observations: Determination of Cosmological Parameters,''
Astrophys.\ J.\ Suppl.\  {\bf 148}, 175 (2003), astro-ph/0302209.
}
\lref\BennettBZ{ C.~L.~Bennett {\it et al.}, ``First Year
Wilkinson Microwave Anisotropy Probe (WMAP) Observations:
Preliminary Maps and Basic Results,'' Astrophys.\ J.\ Suppl.\ {\bf
148}, 1 (2003), astro-ph/0302207.
}

\lref\UngarelliQB{ C.~Ungarelli, P.~Corasaniti, R.~A.~Mercer and
A.~Vecchio, ``Gravitational waves, inflation and the cosmic
microwave background: Towards testing the slow-roll paradigm,''
Class.\ Quant.\ Grav.\  {\bf 22}, S955 (2005), astro-ph/0504294.
}

\lref\FreeseRB{
   K.~Freese, J.~A.~Frieman and A.~V.~Olinto,
   ``Natural Inflation With Pseudo - Nambu-Goldstone Bosons,''
   Phys.\ Rev.\ Lett.\  {\bf 65}, 3233 (1990).
}

\lref\AdamsBN{
   F.~C.~Adams, J.~R.~Bond, K.~Freese, J.~A.~Frieman and A.~V.~Olinto,
   ``Natural inflation: Particle physics models, power law spectra
for large scale structure, and constraints from COBE,''
   Phys.\ Rev.\ D {\bf 47}, 426 (1993), hep-ph/9207245.
}


\lref\DodelsonHR{
   S.~Dodelson, W.~H.~Kinney and E.~W.~Kolb,
``Cosmic microwave background measurements can discriminate among
inflation models,''
   Phys.\ Rev.\ D {\bf 56}, 3207 (1997), astro-ph/9702166.
}

\lref\Lyth{
  D.~H.~Lyth and A.~Riotto,
  ``Particle physics models of inflation and the cosmological density
  perturbation,''
  Phys.\ Rept.\  {\bf 314}, 1 (1999), hep-ph/9807278.
}

\lref\BoyleSE{ L.~A.~Boyle and P.~J.~Steinhardt, ``Probing the
early universe with inflationary gravitational waves,''
astro-ph/0512014.
}

\lref\ByrnesTH{
  C.~T.~Byrnes and D.~Wands,
  ``Scale-invariant Perturbations from Chaotic Inflation,'' astro-ph/0512195.
}

\lref\AdamsVC{ J.~A.~Adams, B.~Cresswell and R.~Easther,
``Inflationary perturbations from a potential with a step,''
Phys.\ Rev.\ D {\bf 64}, 123514 (2001), astro-ph/0102236.
}

\lref\RigopoulosUS{ G.~I.~Rigopoulos, E.~P.~S.~Shellard and
B.~W.~van Tent, ``Quantitative bispectra from multifield
inflation,'' astro-ph/0511041.
}

\lref\SeljakXH{
U.~Seljak {\it et al.},
 ``Cosmological parameter analysis including SDSS Ly-alpha forest and  galaxy
bias: Constraints on the primordial spectrum of fluctuations,
neutrino mass, and dark energy,'' Phys.\ Rev.\ D {\bf 71}, 103515
(2005), astro-ph/0407372.
}


\noblackbox \Title{\vbox{\baselineskip12pt\hbox{hep-th/0512102}
\hbox{PUPT-2184} }}
 {\vbox{ {\centerline{Random Matrices and the Spectrum of N-flation }
}}} \centerline{Richard
Easther$^{a}$\foot{richard.easther@yale.edu} and Liam
McAllister$^{b}$\foot{lmcallis@princeton.edu}}
\bigskip {\it \centerline{$^{a}$Department of Physics, Yale
University, New Haven CT 06520}
\smallskip
\centerline{$^{b}$Department of Physics, Princeton University,
Princeton NJ 08540}}
\bigskip \noindent

N-flation is a promising embedding of inflation in string theory
in which many string axions combine to drive inflation \nf. We
characterize the dynamics of a general N-flation model with
non-degenerate axion masses.  Although the precise mass of a
single axion depends on compactification details in a complicated
way, the distribution of masses can be computed with very limited
knowledge of microscopics: the shape of the mass distribution is
an emergent property.  We use random matrix theory to show that a
typical N-flation model has a spectrum of masses distributed
according to the Mar\v{c}enko-Pastur law.  This distribution
depends on a single parameter, the number of axions divided by the
dimension of the moduli space.  We use this result to describe the
inflationary dynamics and phenomenology of a general N-flation
model. We produce an ensemble of models and use numerical
integration to track the axions' evolution and the resulting
scalar power spectrum.  For realistic initial conditions, the
power spectrum is considerably more red than in single-field
$m^2\phi^2$ inflation.  We conclude that random matrix models of
N-flation are surprisingly tractable and have a rich phenomenology
that differs in testable ways from that of single-field
$m^2\phi^2$ inflation.

\Date{December 2005}

\listtoc
\writetoc

\newsec{Introduction}

Inflationary model-building in string theory
\refs{\stringinflation,\shifty} is advancing rapidly, impelled by
the strong and growing evidence for inflation, and facilitated by
a series of technical advances \refs{\GKP,\kklt,\technicaladvance}
in string theory. Among the inflationary models that have remained
difficult to embed in string theory are those known as large-field
models, in which the inflaton vacuum expectation value (vev)
varies by much more than one Planck mass during inflation
\Dodelson. This class includes what are arguably the simplest
inflationary potentials, those quadratic \chaotic\ or quartic in
the inflaton.

The controllability of these models is a subject of  debate
\liddlelyth, and depends strongly on   the relation between
particle physics and gravity near the Planck scale.  Many authors
take the view that configurations with trans-Planckian vevs are
controllable provided that the energy density is well below the
Planck scale.  Because large-field inflation requires a very
weakly coupled field, potentials such as $\lambda \phi^4$ and $m^2
\phi^2$ automatically have safely sub-Planckian energy densities.
This argument is reliable if we restrict our attention to quantum
corrections that involve higher-order terms in the spacetime
curvature ${\cal{R}}$, since ${\cal{R}} \ll M_{p}^2$ during the
portion of inflation that is observationally relevant.   In a
model that does not unify gravity with the other forces we might
ignore gravitationally induced corrections to the particle physics
effective potential, and only worry about quantum corrections to
the spacetime background.  In this context, trans-Planckian vevs
can indeed be benign.

However, gravitationally-suppressed couplings between the fields
that contribute to $T_{\mu\nu}$ generically appear in models that
unify gravity and particle physics, such as supergravity and
superstring models.  The problem is that loop corrections
generically become large when vevs, instead of energy densities,
exceed the Planck scale; for example, it is very difficult to
forbid corrections to the K\"ahler potential that are controlled
by field vevs in Planck units.  Experience with supergravity and
superstring models strongly suggests that Planck scale vevs
usually imply large corrections to the inflaton effective
potential.  This observation has been used to put bounds on
inflationary observables, such as the tensor power spectrum
\Lyth.\foot{There are a few exceptions to this rule
\refs{\Kawasaki,\Linde}. }

Our view is that because trans-Planckian vevs are always
dangerous, if not necessarily fatal, in string theory and
supergravity models, it is worthwhile to construct specific models
in which these problems can be overcome.  In the best case,
understanding trans-Planckian vevs in a concrete string theory
construction would give a useful toy model for understanding
general quantum gravity corrections to large-field models.

Dimopoulos, Kachru, McGreevy, and Wacker \nf\ have proposed a
clever solution to this set of problems.\foot{For earlier work in
this direction, see \Kanti.  For an application of these ideas to
quintessence, see \KaloperAJ.} Combining $N \gg 1$ fields with
sub-Planckian displacements into a single effective inflaton, they
obtained $m^2\phi^2$ inflation without having to build a
trans-Planckian single-field displacement.  The resulting model,
which they called N-flation, is an example of {\it{assisted}}
\refs{\assist,\Kanti,\allassist,\kaloperliddle,\Jokinen}
inflation. They further explained that N-flation can be embedded
in string theory by choosing the $N$ fields to be axions.  Axions
are typically present in large numbers in string
compactifications, and even when all other moduli are stabilized,
the axion potentials remain rather flat as a consequence of
well-known nonrenormalization theorems \dine.

The axions are pseudo-Nambu-Goldstone bosons associated with
nonperturbatively-broken shift symmetries.  Hence, N-flation may
be categorized as an approximately shift-symmetric, closed string,
large-field model.\foot{Natural inflation
\refs{\FreeseRB,\AdamsBN}, in contrast, is a small-field model
driven by a single axion.} Shift-symmetric D-brane configurations
have been suggested \shifty\ as a natural origin of flat inflaton
potentials in string theory; however, the effects that break such
open string shift symmetries \noshift\ are often too strong to
leave a suitably flat potential. (See also \littleinflaton\ for a
general explanation of the role of approximate shift symmetries in
preserving flat potentials.)

When the individual axion displacements are small in Planck units,
the leading term in the potential is a mass matrix that connects
the $N$ axion fields.  Choosing a basis in which the kinetic terms
are canonical and the mass terms are diagonal, the data of an
N-flation model becomes simply a list of $N$ masses.  The authors
of \nf\ focused their attention on the special case in which the
masses are identical, and showed that in that case the
observational predictions are precisely those of $m^2\phi^2$
chaotic inflation. They argued that inflation still occurs in the
general case with non-degenerate masses, with qualitatively
similar results.  The dynamics is somewhat more complicated,
however, as the most massive axions relax first to their minima,
diminishing the Hubble friction and causing the lighter axions to
roll more rapidly.

The purpose of this note is to give a quantitative treatment of
this general case of N-flation, in which all the masses are
distinct. Two natural questions arise in this context. First, what
are the dynamics and the observational predictions of an N-flation
model with a given set of masses $m_{i}$?  In particular, how does
a spread in masses affect the spectral index $n_{s}$? Second, what
spectrum of masses should one expect from microscopic
considerations? The mass of each axion depends on a formidable
array of compactification details.  In the context of KKLT moduli
stabilization  \kklt, these details include some as-yet uncomputed
factors, such as fluctuation determinants for Euclidean D3-branes,
in addition to better-known but highly unwieldy expressions
involving the fluxes and the K\"ahler potential.  Computing the
precise masses of $N$ axions in a nontrivial, stabilized
compactification would therefore appear to be extremely difficult.
Might we nevertheless be able to deduce something about the
typical spacing of axion masses?

Surprisingly, we find an essentially universal result for the
axion masses in N-flation: the probability distribution for the
masses-squared is given by an analytic formula known as the
Mar\v{c}enko-Pastur law.  The shape of this curve depends on a
single parameter $\beta$ determined by the dimensions of the
K\"ahler and complex structure moduli spaces.  We describe this
result as universal because it does not depend on specific details
of the compactification, such as the intersection numbers, the
choice of fluxes, or the location in moduli space. The result is
also insensitive to superpotential corrections such as threshold
corrections and instanton determinants.  The shape of the mass
distribution depends only on the basic structure of the mass
matrix, which is specified by the supergravity potential. (The
overall mass scale is not determined {\it{a priori}}; as in \nf,
we fix it by requiring the observed amplitude of density
perturbations.)

This welcome situation arises because of the size (typically
larger than $200 \times 200$) of the matrices involved: the
distribution of the eigenvalues of random matrices is most easily
characterized when the matrices are large. In this way, the
simplifications enjoyed by large random matrices
\refs{\mehta,\wign,\wignrev,\bai} help N-flation to generate a
prediction for the scalar spectral index which is substantially
(though not completely) independent of compactification details.
This is a striking example of a problem that is tractable because
of, rather than in spite of, the high dimensionality of the scalar
field space in string theory.

Equipped with the Mar\v{c}enko-Pastur law for the mass
distribution, we numerically integrate the equations of motion for
a large sample of realizations of N-flation.  We find that for a
range of initial conditions and for most values of $\beta$, the
scalar power spectrum is typically more red ({\it{i.e.}} farther
from scale-invariant) than in the case where all the masses are
identical.

In addition to developing a random matrix theory calculation of
the axion mass spectrum, we treat the inflationary dynamics and
perturbation spectrum produced by N-flation.  (For related earlier
work, see \refs{\bartjanphd,\Alabidi}.  Additionally, Kaloper and
Liddle \kaloperliddle\ have considered inflation driven by a tower
of Kaluza-Klein fields with quadratic potentials. However, this
mass distribution is very different from the Mar\v{c}enko-Pastur
form relevant in our case of string axions.) It turns out that
while there are very general results for the perturbation spectrum
produced by multi-field inflation ({\it{e.g.}} \Sasaki), the
spectrum of N-flation is surprisingly complex and has several
previously unsuspected features. In this paper, we work only with
the slow-roll approximation, and the results here represent an
initial survey rather than a comprehensive analysis of the
subject. We would like to emphasize that the dynamics and the
resulting inflationary power spectrum do {\it{not}} become
independent of the initial conditions at late times
\kaloperliddle.  Thus, even if one had a microscopically
computable N-flation model, some of the predictions would depend
on the initial conditions.  For this reason, we present our
results as functions of the initial conditions, illustrating the
simplifications that arise in a few special cases.  A theory of
the initial conditions, which we do not address in this paper,
would still be necessary for a complete understanding of this
class of models.

Readers whose main interests are the inflationary dynamics and the
predictions of the model, rather than the problem of the axion
mass spectrum in a string compactification, could begin with
\S2.1, examine the mass spectrum of \S4.2, and then proceed from
\S6.

\newsec{The N-flation Model }

\subsec{Basic Form of the Model}
 Here we briefly recall the structure of
the model proposed by Dimopoulos {\it{et al.}}\nf.  Consider $N$
axions $\varphi_{i}$ whose periodic potentials \eqn\vofin{
V(\varphi_{i}) =
\Lambda_{i}^{4}~\Bigl(1-{\rm{cos}}\Bigl({\varphi_{i}\over{f_{i}}}\Bigr)\Bigr)}
are independent and arise solely from nonperturbative effects.  As
we will explain, we omit multi-instanton corrections to this
potential.  The scale $\Lambda_{i}$ is determined by dimensional
transmutation and can be hierarchically small compared to the
ultraviolet cutoff.  We find it convenient to absorb the axion
decay constants $f_{i}$ via \eqn\dec{\phi_{i} \equiv
{\varphi_{i}\over{f_{i}}},} and we take units where $M_{p}=1$. For
small axion displacements $\phi_{i} \ll 1$ , the potential may be
Taylor expanded, and the Lagrangian is\foot{For the moment we have
neglected the important possibility of nontrivial axion kinetic
terms, but we will come to a complete treatment of this point.}
\eqn\lgs{ L = {1\over{2}}\nabla_{\mu}\phi_{i}\nabla^{\mu}\phi_{i}
- {1\over{2}}m_{i}^2\phi_{i}^2}

If the masses $m_{i}$ are very nearly equal and the axions are
initially displaced by a common amount $\bar{\phi}$, they will
roll in unison toward their joint origin.  For all practical
purposes, this is equivalent to the motion of a single field
$\Phi$ with a displacement $\sqrt{N}\bar{\phi}$.  It follows that
even if each axion is displaced by a distance in field space small
compared to the Planck mass, $\Phi$ can satisfy the conditions for
slow-roll inflation.  In this sense, N-flation realizes
$m^2\phi^2$ inflation in a very well-controlled string theory
setting.

One should ask how often this simple potential involving $N$
decoupled axions with canonical kinetic terms and identical masses
can arise in a string compactification.  We would expect that a
generic model with $N$ decoupled axions will have a nontrivial
spectrum of masses.  Our goal is to determine exactly which sorts
of mass spectra are possible in string compactifications.
Surprisingly, we will find that a single characteristic spectrum
emerges, regardless of microscopic details.

\subsec{N-flation in a Stabilized Vacuum}

An important requirement for a realistic inflationary model in
string theory is the stabilization of compactification moduli. For
this reason, it is important to verify in detail that the
particular classes of axion potentials needed for N-flation do
arise in a class of stabilized string compactifications.

The moduli are typically paired with axions in complex
combinations. In the type IIB theory, for example, the K\"ahler
modulus $\rho_{i}$ measuring the volume of a given four-cycle
$B_{4}^{(i)}$ combines with the axion \eqn\axtwob{ \phi_{i} =
\int_{B_{4}^{(i)}}C_{4}} as $\rho_{i} - i \phi_{i}$.  (We have
chosen a convenient sign.)

There is a simple relationship between the moduli-stabilizing
potentials and the axion potentials in the resulting stabilized
vacuum.  For definiteness, we will focus on moduli stabilization
in type IIB string theory by the method of KKLT \kklt, in which
all moduli are stabilized by the combination of fluxes \GKP\ and a
nonperturbative superpotential.  In the type IIA vacua of \twoa,
most fields receive masses from fluxes, but the lightest axions
are lifted by nonperturbative effects; for this reason, we expect
rather similar qualitative features in this context.\foot{This
conclusion was reached in discussions with A. Krause and A.
Mazumdar.}  We leave the properties of N-flation in more general
stabilized vacua as an interesting subject for future work.

The KKLT superpotential is \eqn\wis{ W = W_{0}(S,\chi_{a}) +
\sum_{i} A_{i}(\chi_{a}) e^{- a_{i}\tau_{i} } .} Here \eqn\wflux{
W_{0} = \int G_{3} \wedge \Omega } is the flux-induced
superpotential, which depends on the dilaton $S$ and on the
$h_{2,1}$ complex structure moduli $\chi_{a}$.  The sum of
nonperturbative superpotential terms runs over all the K\"ahler
moduli $\tau_{i} \equiv \rho_{i}- i \phi_{i} ~~(i=1 \ldots
h_{1,1})$.  The real parts $\rho_{i}$ measure the volumes of
four-cycles in the Calabi-Yau, and the imaginary parts $\phi_{i}$
are the axions arising from the four-form potential $C_{4}$.  We
will work at volumes large enough to ensure that neglecting the
multi-instanton corrections to \wis\ is consistent (see \S9.1).

We are assuming that there is one nonperturbative effect for each
independent four-cycle. For each such cycle, there are two
possible sources of a nonperturbative superpotential: Euclidean
D3-branes or strong infrared dynamics in a gauge theory, for which
gaugino condensation on a stack of $N>1$ D7-branes is the simplest
example.  The constants $a_{i}$ are either $2\pi$, in the case of
Euclidean D3-branes, or $2\pi/N$ for gaugino condensation. For
simplicity we will assume that Euclidean D3-branes provide all the
nonperturbative superpotential terms; relaxing this assumption is
straightforward and does not affect our results.  The prefactors
$A_{i}(\chi_{a})$, although knowable in principle, are unknown in
practice.  They stem from determinants of fluctuations on the
D3-brane-instanton worldvolume.

Finally, the K\"ahler potential is given, to leading order, by
\eqn\kis{ K = - 2 ~{\rm{log}}\Bigl({\cal{V}}\Bigr)} where
${\cal{V}}$ is the Calabi-Yau volume, amounting to a highly
nontrivial combination of the various K\"ahler moduli $\rho_i$.
(When there is only a single K\"ahler modulus, ${\cal{V}} \propto
\rho^{3/2}$.)

The only feature of the KKLT scenario that will be important for
our discussion is that the superpotential is a sum of an
axion-independent term \GVW\ and $N$ nonperturbative terms in
which the axions appear as phases. This would still be true in an
appropriate generalization of the KKLT method to other string
theories.  When this is the case, the axions potentials arise
exclusively from the same nonperturbative effect that stabilizes
the K\"ahler moduli with which they are paired.  Our qualitative
conclusions do not even depend on this basic property, but for the
purpose of presenting a definite computation, we do assume the
KKLT form except where noted.

\subsec{F-flatness Conditions }

We are interested in a scenario in which the axions' displacement
from their joint minimum is the dominant source of positive energy
during inflation.  Once N-flation ends and the axions have relaxed
to this minimum, supersymmetry may still be broken by some
additional effect.  However, this effect should be subleading, and
ideally constant, during inflation, lest it ruin the useful
properties of the axion-induced inflationary potential.

The F-terms $D_{A}W$ of the various moduli $S,\chi_{a},\rho_{i}$
depend on the superpotential, and through it, on each of the
axions.  As the axions vary, the F-terms change, so the
inflationary trajectory cannot be F-flat.  However, we will
substantially simplify our analysis by approximating the final
vacuum as F-flat.  This is an excellent approximation for
phenomenologically viable models with low-energy supersymmetry:
the F-term energy at late times can be at most of order the
intermediate scale, and hence will be highly suppressed relative
to the inflationary energy of the axions.\foot{We thank S. Kachru
for helpful discussions of this point.}  Correspondingly, we will
refer to configurations as F-flat even if they satisfy only the
weaker condition that the F-term energy is very small compared to
the energy scales of interest for inflation.  We could also
imagine that the dominant supersymmetry-breaking in the final
state is accomplished by a D-term effect such as an anti-D3-brane.
In any event, we require that as the axions roll to their minima,
the configuration relaxes to a KKLT vacuum, \eqn\change{
V(\phi_{f},S,\chi,\rho)=V_{KKLT}(S,\chi,\rho)} with $\phi_{f}$ the
value the axions take at their minima. To summarize, then, our
picture is that the axions are displaced from the minima they take
in a supersymmetric $AdS_{4}$ vacuum, and their potentials are
unaffected by any additional, subleading supersymmetry-breaking
effects.

We assume that all other moduli are fixed during inflation.  By
construction, all the moduli have nonzero masses in the KKLT
vacuum, but we are requiring in addition that these masses are
somewhat larger than the axion masses.  A condition of this nature
is essential to prevent unfixed moduli, whose potentials are not
protected against corrections, from spoiling the properties of an
inflationary epoch in which only the axions are dynamical.  At
first sight it may appear implausible that a modulus field
$\rho_{i}$ would be firmly stabilized while its axion partner
$\phi_{i}$ from the same chiral multiplet remains somewhat mobile.
The important difference is that the $\rho_{i}$ appear in the
K\"ahler potential, whereas the $\phi_{i}$ do not; as a result,
when the KKLT potential is expressed in terms of
canonically-normalized fields, it is steeper in the $\rho_{i}$
directions than in the $\phi_{i}$ directions. Equivalently, the
difference is that the $\phi_{i}$ enjoy the protection of an
approximate shift symmetry, but the $\rho_{i}$, like the other
moduli, do not.  During inflation the unprotected moduli will
typically acquire masses $ m \gtrsim H$, while the axion masses
will remain substantially smaller.\foot{See {\it{e.g.}} \banks\
for a discussion of related issues.}

To find the supersymmetric $AdS_{4}$ vacuum, we solve \eqn\fterm{
0 = D_{A}W \equiv
\partial_{A}W + \left(\partial_{A}K\right)W }
where $A$ runs over the dilaton $S$, the complex structure moduli
$\chi_{a}$, and the K\"ahler moduli $\rho_{i}$.  The F-terms
\fterm\ are axion-dependent, and cannot vanish unless $\phi=0$ or
$\pi$. However, the F-terms also depend on the other moduli
$S,\chi,\rho$. As explained above, while the axions are supporting
inflation, the other moduli must remain fixed. Together with
\change, this implies that the full axion potential is
\eqn\axpot{V(\phi)=V(\phi,\rho_{i}^{(0)},\chi_{i}^{(0)},S^{(0)})}
{\it{i.e.}} the supergravity potential evaluated on a background
configuration in which $\rho_{i},\chi_{a},S$ take the same values
they take at the global minimum.  Although the F-terms do not
vanish along the axion trajectory, but only at the endpoints, one
must still enforce \eqn\chisect{ D_{A}W(\phi_{i}=0)=0}  This
amounts to setting the axions to zero in each of the F-terms and
using the residual relations to constrain $\rho_{i},\chi_{a}, S$.
These constraints will provide very useful clues about the
structure of the axion potential.

\newsec{Mass Terms in N-flation}

We will now work out in some detail the potential arising from a
KKLT realization of N-flation.

\subsec{The Quadratic Potential}

Recall that the axion kinetic terms are identical in structure to
those of the K\"ahler moduli with which they are paired, so that
\eqn\axk{ {\cal{L}}_{axion} = {1\over{2}}M_{p}^2
K_{ij}\nabla_{\mu}\phi^{i}\nabla^{\mu}\phi^{j}-V} The supergravity
potential is \eqn\massis{ V = {\rm{exp}}
\Bigl({K\over{M_{p}^2}}\Bigr) \Bigl(K^{AB}D_{A}W D_{\bar
B}\overline{W} -3{|W|^2\over{M_{p}^2}} \Bigr)} where, as before,
$A,B$ run over the dilaton $S$, the complex structure moduli
$\chi_a$, and the K\"ahler moduli $\rho_i$.  (In this section we
find it convenient to retain explicit factors of $M_{p}$, which we
have set to unity elsewhere.)

In the KKLT superpotential \wis, each nonperturbative factor
depends on the associated axion as \eqn\wnpis{ W_i = A_{i}
e^{-2\pi \rho_i} e^{2\pi i\phi_i} \equiv C_{i} e^{2\pi i\phi_i} }
We have chosen the notation $C_{i}$ to stress the fact that the
moduli $\chi_{a},\rho_{i}$ which appear in $C_{i}$ are {\it{not}}
dynamical variables in the scenario of interest: by \fterm, these
fields are fixed to the values they take in the supersymmetric
minimum. Thus, the $C_i$ are constants as far as the inflationary
dynamics is concerned.  Because the axion $\phi_{i}$ appears only
in $W_{i}$, and neither in the K\"ahler potential nor elsewhere in
the superpotential, every factor of $e^{2\pi i\phi_{i}}$ in
\massis\ will appear multiplied by the constant $C_{i}$.

Inserting \wis\ in \massis, performing a Taylor expansion around
the origin $\phi_i=0$, and using the F-flatness conditions
$D_{A}W|_{\phi_{i}=0}=0$, we arrive at \eqn\vmassmat{ V = (2\pi)^2
\hat{M}_{ij}\phi^i \phi^j  + {\cal{O}}(\phi^3)} where
\eqn\massmatrix{ \hat{M}_{ij} = {1\over{M_{p}^2}}e^{K}
\Bigl(K^{AB}D_{A}C_{i} D_{B}C_{j}-3 C_{i}C_{j} \Bigr)}

The kinetic terms \axk\ and the mass matrix \vmassmat\ are
evidently {\it{not}} diagonal in general.  Thus, we find that the
potential does {\it{not}} automatically take on the uncoupled form
required for N-flation. That is, the mass matrix is typically not
diagonal in the the basis in which the superpotential is simple,
with a single axion appearing in each nonperturbative
term.\foot{We thank S. Kachru and J. Wacker for useful discussions
of this point.} Moreover, the off-diagonal terms appear to be of
the same order as the diagonal ones.  We show in Appendix A that
for essentially statistical (rather than directly physical)
reasons, the cross-couplings in $\hat{M}_{ij}$ are actually
suppressed relative to the diagonal terms. Even so, keeping track
of these cross-couplings will provide a key insight into the mass
spectrum.

At this stage, it is convenient to make a change of basis that
renders the axion kinetic terms canonical.  Define $O$ to be the
matrix that diagonalizes the metric $K_{ij}$ on the K\"ahler
moduli space, \eqn\diagz{ O_{i}^{~k}K_{kl}{O_{j}^{~l}}=
\kappa_{i}^2 \delta_{ij}} with $\kappa_{i}^2$ the positive
eigenvalues. The eigenvalues $\kappa_{i}$ are related to the axion
decay constants\foot{According to \BanksSX, the $\kappa_{i}$ are
at most of order unity; see also \wittalk.} through \eqn\kapf{
\kappa_{i} = {f_{i}\over{M_{p}}}} Next, rescale the $\phi_{i}$ to
absorb the factors $\kappa_{i}$, so that the kinetic terms are
canonically normalized and $K^{ij}$ is the $N \times N$ identity
matrix. Finally, redefine the complex structure moduli and the
dilaton so that $K^{ab}$, $a,b = 1, \ldots P$ is the $P \times P$
identity matrix, and return to units in which $M_{p}=1$.

This leads to the Lagrangian density \eqn\diagk{{\cal{L}}=
{1\over{2}}\partial_{\mu}\phi_{i}\partial^{\mu}\phi^{i}-
M_{ij}\phi^{i}\phi^{j} } where
\eqn\rotmassmatrix{ M_{ij} = (2\pi)^{2}{{e^K}\over{f_{i}f_{j}}}
O_{i}^{~k}\Bigl(D_{A}C_{k}D^{A}C_{l}-3 C_{k}C_{l} \Bigr)
O_{j}^{~l}}

Provided that the potential takes this purely quadratic form,
{\it{i.e.}} neglecting the higher-order terms omitted in
\vmassmat, there exists a further orthogonal rotation that
diagonalizes $M_{ij}$.  The result is that all the data specifying
an N-flation model can be stored in the list of eigenvalues of
\rotmassmatrix, which are the masses-squared $m^{2}_{i}$ of $N$
canonically-normalized, uncoupled axions.  Our goal in the
remainder of the paper is to characterize the statistical
properties of these eigenvalues, based on the structure of
\rotmassmatrix.

\subsec{The Complete Potential}

In the previous section we focused on the quadratic potential
\vmassmat.  However, terms of higher order in the axions must be
retained when some of the initial dimensionless axion
displacements $\phi_{i}^{(0)}$ are of order one.  When the
quadratic approximation to the axion potential is invalid, it is
generally impossible to diagonalize the potential to remove
cross-coupling terms.

Because large initial displacements are advantageous for
inflation, there is a competition between obtaining enough
inflation and having a controllable expansion of the full
potential \massis\ as a potential for uncoupled fields.  For that
reason, in this section we will present the full, unexpanded
potential \massis\ and sketch the resulting axion dynamics.

To do this, we reorganize \massis\ and use the F-flatness
conditions \chisect.  The result from the F-term piece\foot{The
terms from $-3|W|^2$ have an analogous form but complicate the
formulas substantially.  Because these terms are unimportant for
our considerations, we omit them here.} of the potential is
\eqn\manycos{V=
\sum_{i,j}\Bigl(e^{K}K^{AB}D_{A}C_{i}D_{B}C_{j}\Bigr)\Bigl(1+{\rm{cos}}(2\pi\phi_i-2\pi\phi_j)-2~{\rm{cos}}(2\pi\phi_i)\Bigr)}
Taylor expansion of this potential leads immediately to the F-term
part of \vmassmat.

The potential \manycos\ can be usefully rewritten in a more
general form \eqn\rewr{ V = V_{0} +
\sum_{i}\alpha_{i}{\rm{cos}}(2\pi\phi_i)+\sum_{i,j}\beta_{ij}{\rm{cos}}(2\pi\phi_i-2\pi\phi_j)}
from which we can recover \manycos\ if we take the obvious
definitions of the axion-independent term $V_{0}$ and of the
coefficients $\alpha_{i}, \beta_{ij}$.

The simplest sort of N-axion potential would have been a sum of
cosine potentials, each involving a single field, with negligible
cross-couplings, {\it{i.e.}} $\beta_{ij}=0$.  We have just seen
that such a simple potential never governs deviations from an
approximately F-flat KKLT minimum.  While we do not claim that
uncoupled N-axion potentials are impossible in string
constructions, we do expect that they are rather rare in general.
It would be very interesting to find exceptions to this rule.

Requiring that the $\beta_{ij}$ are negligibly small is also a
very significant constraint on the general form \rewr. The problem
is that there are $N^2$ cross-coupling terms but only $N$
single-field terms, so cross-couplings which appear small could
still be important in large numbers.  (However, as explained in
\nf, the fact that each field appears in only a small fraction of
the couplings suggests that assisted inflation is still possible.
See the Appendices for a more detailed discussion of these
issues.)

In our particular case \manycos, it is apparent that the net
effect of the terms coupling two axions is not subleading to the
effect of the terms involving a single axion.  For example, when
all the axions are near the origin $\phi_{i}=0$, the two classes
contribute amounts of energy that differ only by a factor of two.
If instead the axions have a common vev $\bar{\phi} \sim
{\pi\over{2}}$, the cross-terms give the dominant contribution to
the potential. Moreover, the energy from these cross-terms is
positive and acts to repel the axions from the same point of their
configuration space, substantially complicating the motion of the
axions.  The identification of a single direction for collective
motion (such the inflaton $\rho$ of \nf) could be challenging.  We
will leave the full dynamics of the N-axion potential \rewr\ as an
interesting problem for the future.

Finally, it is important to observe that diagonalizing the mass
matrix is not equivalent to sending $\beta_{ij} \to 0$: the
presence of cross-couplings ensures that the eigenvalues of
$M_{ij}$ are different from the $\alpha_{i}$. The correct
procedure, when the quadratic approximation is applicable, is to
rotate away the cross-couplings, not to omit them.  Surprisingly,
this simplifies the problem of understanding the mass spectrum: we
will see that although the collections $\{\alpha_{i}\}$ vary
considerably from one compactification to another, the statistical
properties of the spectrum of eigenvalues $\{m_{i}^{2}\}$ of
$M_{ij}$ are much less variable.

\newsec{Random Matrix Treatment of the Mass Terms}

In order to understand which classes of N-flation models are most
likely, we must understand the properties of the mass matrix
$M_{ij}$, \rotmassmatrix.  However, the complete expression is
prohibitively complicated, particularly because it depends on the
unknown quantities $D_{A}C_{i}$.  For this reason, we resort to
characterizing the expected properties of the {\it{ensemble}}
$\Omega_{M}$ of mass matrices one expects in an ensemble of
N-flation models.  More specifically, we seek the probability
distribution function $p(m^{2})$ for the eigenvalues $m^{2}_{i}$
of the mass matrix, a function which we call the mass spectrum for
short.

The logic behind this step is that, even without knowing the
individual terms $D_{A}C_{i}$, we can learn something about the
eigenvalues of the mass matrix by searching for properties of the
distribution of eigenvalues of \rotmassmatrix\ that are
independent of the particular set of values taken on by the
$D_{A}C_{i}$ in any given realization.  This approach is effective
because the mass matrices are so large that fluctuations away from
the mean behavior are unlikely. Thus, knowledge about average
elements of $\Omega_{M}$ provides an excellent guide to the
properties of nearly all the members of the ensemble.  We will
therefore restrict ourselves to characterizing $\Omega_{M}$ as a
whole.

Moreover, although we will be able to establish some facts about
the statistics of the $D_{A}C_{i}$ themselves,\foot{More
sophisticated treatments of the statistics of flux vacua appear in
\douglas.} it is possible, and advantageous, to find properties of
the eigenvalue spectrum that are substantially independent not
just of the particular values attained by the $D_{A}C_{i}$, but
even of the {\it{distribution}} $\Omega_{D}$ of these values in an
ensemble of models.  Because the matrices in question are
moderately large, simplifications of the sort that often arise at
large $N$ in random matrix theory turn out to simplify the
problem.  A particular result \bai\ which provides a strong
motivation for our approach is this: the behavior of the
eigenvalue spectra of $N \times N$ random matrices whose entries
have appropriately bounded moments but an otherwise arbitrary
distribution $\Omega_{D}$, is identical, in the $N \to \infty$
limit, to the spectrum in the simplest case in which the entries
are independent and identically distributed (henceforth i.i.d.)
and $\Omega_{D}$ is a Gaussian distribution with mean zero. This
is an extremely powerful statement, and although we will not
appeal to it directly, it gives an underlying reason that one
might have expected universal behavior in our system.

\subsec{Statistical Model for the Covariant Derivatives}

In the basis in which the kinetic terms are canonical, the mass
matrix takes the form \rotmassmatrix.  The mass terms arising from
$-3|W|^2$ are subleading, by a factor of the inverse volume
${1\over{\rho}}$, to those coming from the F-terms. For this
reason, we neglect the former in the analytical discussions that
follow.  In \S5.2 we will present the results of a Monte Carlo
simulation of the full structure \rotmassmatrix\ to show that this
approximation is a sound one.  There we will find that even
{\it{without}} including the suppression by ${1\over{\rho}}$, the
terms from $-3|W|^2$ have a negligible effect on the statistical
properties of the mass matrix.

The remaining quantities of interest are the rotation matrices
$O_{i}^{~j}$, the decay constants $f_{i}$, and, most importantly,
$H_{Ai} \equiv D_{A}W_{i}$, the $(N+P) \times N$ matrix of
K\"ahler-covariant derivatives.

We will now attempt to characterize the statistical properties of
the entries in $H$.  We propose that the entries of $H$ may be
modeled as independent, identically distributed (i.i.d.) variables
with mean $\mu_{H}$ and variance $\sigma_{H}^2$.  We do not assume
that the distribution is Gaussian.

From these assumptions, one can show that in fact $\mu_{H} \ll
\sigma_{H}$.  Observe that the row-average $r_{A} \equiv
{1\over{N}}\sum_{i} H_{Ai} = -{1\over{N}}D_{A}W_{0}$ by the
F-flatness condition.  Granting the i.i.d. assumption stated
above, the Central Limit Theorem implies that the row-averages
$r_{A}$ are drawn from a distribution with mean $\mu_{H}$ and
standard deviation ${\sigma_{H}\over{\sqrt{N}}}$.  The sign of
{\it{every}} $r_{A}$ is therefore the sign of $\mu_{H}$ , unless
$\sigma_{H} \gtrsim \sqrt{N}\mu_{H}$ (so that a fluctuation can
change the sign of one or more of the $R_{A}$.)

However, genericity and microscopic considerations require that
the terms $r_{A} = - {1\over{N}}D_{A}W_{0}$ take on both signs. In
particular, the sign of $ r_{i}= - {1\over{N}}W_{0}\partial_{i}K$
depends on the intersection numbers $C_{ijk}$ via $\partial_{i}K =
-{1\over{{\cal{V}}}}C_{ijk}t^{j}t^{k}.$ N-flation is only possible
in compactifications in which some fraction of the $C_{ijk}$ are
negative (in order to avoid excessive renormalization of Newton's
constant \nf), so the $r_{A}$ cannot all have the same sign.  If
our distribution is to reproduce this basic feature, the mean
value $\mu_{H}$ of the $H_{Ai}$ must be taken to be parametrically
small compared to the standard deviation, \eqn\musup{ \mu_{H}
\lesssim {\sigma_{H}\over{\sqrt{N}}}} For all practical purposes
we may therefore set $\mu_{H}=0$.\foot{Even if an exception to
this result could be found, the addition of a mean $\mu_{H} \neq
0$ would not change the spectrum of masses of the $N-1$ lightest
axions \bai.}

Now define \eqn\risnow{ R_{Ai} = 2\pi e^{K/2}f_{i}^{-1}
O_{i}^{~j}H_{Aj}} so that the leading contribution to the mass
matrix is \eqn\simm{ M_{ij} = R_{iA}R^{A}_{~j}}  Provided that the
rotations $O$ are random, it follows from $\mu_{H}=0$ that the
mean $\mu$ of the $R_{Ai}$ may likewise be approximated by zero.
However, the variance $\sigma^2$ of the $R_{Ai}$ may be different
from $\sigma_{H}^2$.

In summary, we claim that for the purpose of determining the
spectrum, the essential structure of the mass matrix is \simm,
where $R_{Ai}$ is the $(N+P) \times N$ matrix defined in \risnow.
We further propose to approximate the entries of $R$ as i.i.d.
variables with zero mean and some variance $\sigma^2$.  We repeat
that the entries do not necessarily have a Gaussian distribution.
To underscore both this point and the results of \S4.1, we will
construct, in \S5, ensembles of random mass matrices using the
full formula \rotmassmatrix.  For a range of assumptions about the
distributions of the entries of $K^{AB}$ and $H_{Ai}$, we will
find results indistinguishable from those that follow from \simm.

We have not found an exact expression for $\sigma$ in terms of
microscopic parameters, beyond the definition \eqn\defsig{ \sigma
\equiv \langle |R_{Ai}| \rangle } where the average is over all
values of the indices $A,i$.  However, a useful heuristic is
\eqn\heur{ \sigma^2  \sim {{e^K}\over{\langle f^2 \rangle}}
\langle \left(D_{A}C_{i}\right)^2  \rangle} which shows that
$\sigma$ measures the typical size of the F-terms.  The scale of
the total superpotential, and thus of the F-terms, is determined
by size of the flux-induced superpotential $W_{0},$ which in turn
is fixed by a choice of fluxes.  As in other applications of the
KKLT proposal, a fine-tuning of the value of $W_{0}$ is necessary
in order to construct a vacuum at large volume and weak coupling.
Although many moduli are present in our setup, a {\it{single}}
fine-tuning, that of $W_{0},$ suffices to fix the overall scale,
and hence the typical axion mass $\bar{m}$.  In practical terms,
we set the value $\bar{m}$ exactly as in \nf, by requiring that
the density perturbations resulting from inflation have amplitude
consistent with observations.  This implies (see \S8)
\eqn\setover{ \bar{m} \approx 1.5 \times 10^{-5} M_{p}} Let us
point out that \setover\ is not satisfied by an N-axion potential
in a truly {\it{generic}} string compactification, which would
presumably have an average mass of order one in string units.  We
must restrict our attention to the subclass of models (with
fine-tuned fluxes) in which the overall scale does obey \setover.
Of course, this problem is not special to N-flation: in nearly
every model of inflation in field theory or string theory, one
must set the scale of inflation by hand.  In our case, the
necessary fine-tuning happens to be of the form treated by Bousso
and Polchinski \bp.

The distribution of the axion masses around the average scale
\setover\ is the quantity that we will now determine with random
matrix techniques.

\subsec{Axion Masses and the Mar\v{c}enko-Pastur Law}

From the complicated structure of the mass matrix \rotmassmatrix\
we have abstracted one essential property: \eqn\triv{M = R^{T}R }
where $R$ is a $(N+P)\times N$ rectangular matrix whose entries
are i.i.d. with zero mean and variance $\sigma^2$.  We have
therefore reduced the problem of finding the mass spectrum in a
typical N-flation model to that of finding the eigenvalue
probability distribution function $p(m^2)$ for a matrix of the
form \triv.  This important problem\foot{The problem of finding
$p(\lambda)$ for an ensemble of matrices of the form \triv\ is
important for statistical analysis and also has a range of
practical applications, for example in communications engineering
\comm.  Here we denote the eigenvalue by $\lambda$ because the
interpretation of an eigenvalue as a mass-squared is rare in these
other applications.} was solved by Mar\v{c}enko and Pastur in 1967
\mp.

The Mar\v{c}enko-Pastur law for the eigenvalue spectrum is
\eqn\mpis{ p(m^2) = {1\over{2\pi m^2
\beta\sigma^2}}\sqrt{(b-m^2)(m^2-a)}} for $a \le m^2 \le b$, where
we have defined

\eqn\ais{ a = \sigma^2 \Bigl(1-{\sqrt{\beta}}\Bigr)^2} \eqn\bis{ b
= \sigma^2 \Bigl(1+{\sqrt{\beta}}\Bigr)^2} \eqn\betis{\beta =
{{N}\over{N+P}}} The probability density vanishes outside this
range.  In Fig.1 we display the Mar\v{c}enko-Pastur law for a few
illustrative cases.

\fig{The Mar\v{c}enko-Pastur law for $\beta=9/10$ (highest
peak),$~1/3$,$~1/10$ (most localized).}{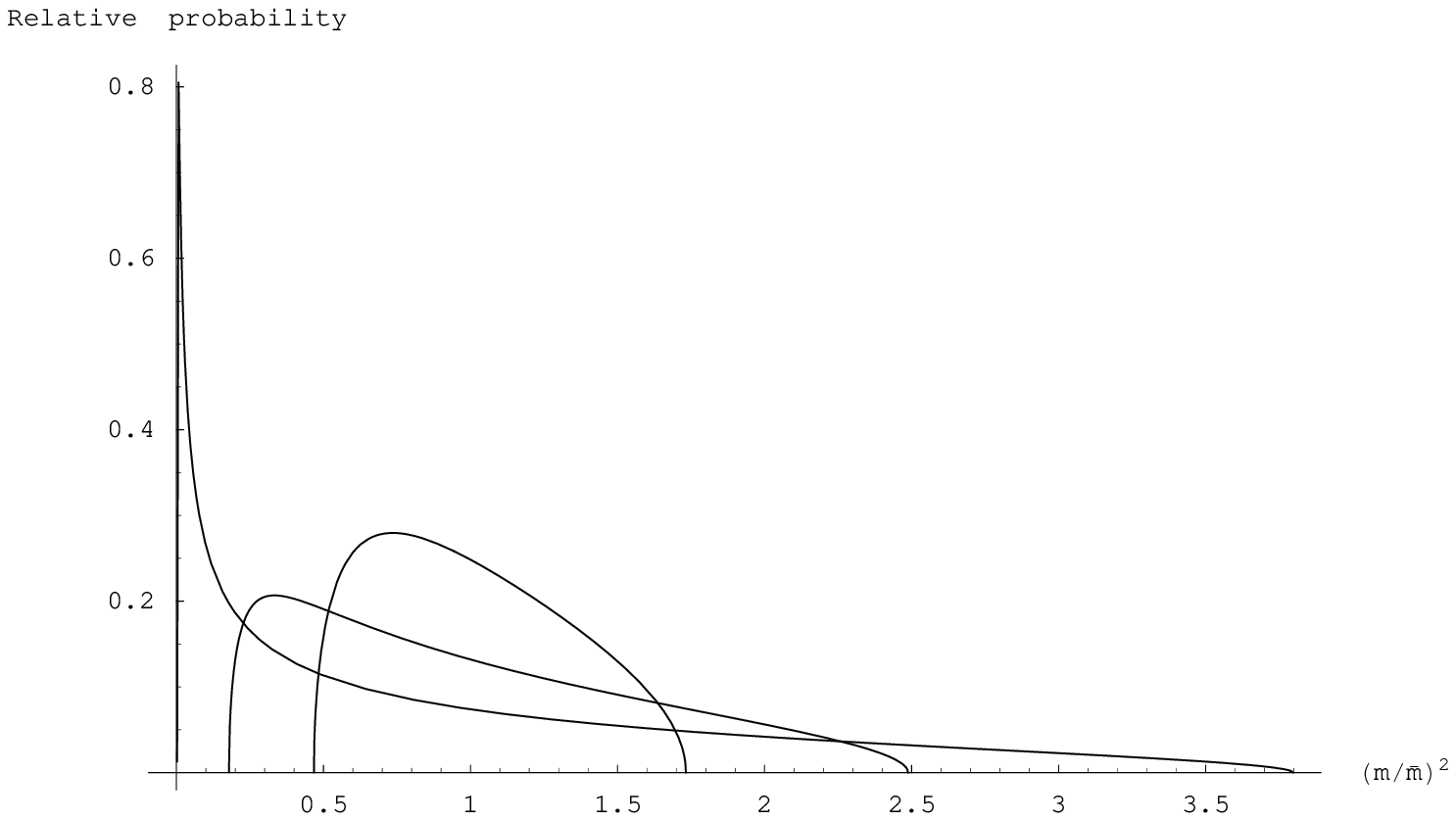}{6truein}

The parameters controlling the shape of the spectrum \mpis\ are
$\beta$ and $\sigma$, where $\beta$ is the ratio of the dimensions
of the rectangular matrix $R$ and $\sigma^2$ is the variance of
the entries of $R_{Ai}$.  In the case of a N-flation in a KKLT
compactification of type IIB string theory, there are $h_{1,1}$
axions, and $h_{1,1}+h_{2,1}+1$ is the total dimension of the
moduli space (K\"ahler, complex structure, and dilaton), so that
\eqn\kkbet{ \beta = {h_{1,1}\over{h_{1,1}+h_{2,1}+1}}}  In more
general cases, for example in other string theories, it may be
possible to identify $\beta$ with a ratio of this form: the number
of (real) axions involved in inflation divided by the total number
of moduli chiral multiplets.

We will not attempt to find the smallest and largest values of
$\beta$ that can arise in a working N-flation model.  In any case,
there is an important, independent constraint on $\beta$.  The
renormalization of Newton's constant described in \nf\ and
reviewed in \S9.1 requires that $N \sim P$, so that $\beta \sim
{1\over{2}}$.  In special (and perhaps fine-tuned) circumstances
this constraint might be relaxed, so we perform the remainder of
our analysis for a general $\beta$. However, we will keep in mind
that models with $\beta \sim {1\over{2}}$ are strongly favored
because they more readily produce an appropriate number of
e-folds.

To determine the value of $\sigma,$ notice from \mpis\ that the
average value of the mass-squared is \eqn\massexp{ \langle m^2
\rangle = \sigma^2 } from which it is evident that $\sigma$ sets
the overall mass scale.  As explained in the previous section, the
only phenomenologically viable N-flation models are those in which
\setover\ is satisfied.  This effectively fixes the value of
$\sigma$ and leaves $\beta$ as the unique parameter specifying an
N-flation model.


\subsec{Some Relations to Other Results in Random Matrix Theory}

The Mar\v{c}enko-Pastur law \mp\ is in some sense analogous to
Wigner's celebrated semicircle law \wign\ for the distribution of
eigenvalues of a random symmetric matrix $ M = H + H^{T},$ where
the entries of the square matrix $H$ are i.i.d. Gaussian variables
with mean zero.  In the same way, $\Omega_{M}$ has some
superficial similarities to the famous Gaussian Orthogonal
Ensemble (GOE), which describes symmetric matrices whose entries
are drawn from normal distributions.

Recall that elements of the GOE \mehta\ may be constructed as $ G
= H + H^{T}$, where the entries of $H$ are i.i.d. Gaussian
variables with zero mean.  The eigenvalues of $G$ obey Wigner's
semicircle law.   Our mass matrix has instead the structure $ M =
H^{T} H$, and substantial correlations among the matrix entries in
$M$ conspire to give the Mar\v{c}enko-Pastur spectrum of
eigenvalues.

The Mar\v{c}enko-Pastur law \mpis\ does approach a semicircle law,
although not one centered at the origin, in the limit $\beta \to
0$.   This limit corresponds to $P/N \to \infty$,  so that the
number of axions is very small compared to the total number of
moduli. Compactification manifolds in which $P \gg N$ and $N \gg
1$ are presumably quite rare, so we find that the
Mar\v{c}enko-Pastur law is quite different from the semicircle law
for realistic values of $\beta$.

In the other limiting case $\beta \to 1$, the Mar\v{c}enko-Pastur
spectrum approaches the {\it{square}} of a Wigner semicircle
spectrum (also known as the quarter-circle law \wignrev); that is,
one can construct the former spectrum by squaring the eigenvalues
appearing in the latter.  Squaring a symmetric but otherwise
general matrix $H$ is one way to construct a matrix which is
positive-definite (as we have demanded that our mass matrix should
be) and has an easily-computed spectrum.  In the limit $\beta \to
1$, in which there are very few chiral multiplets in the theory
except for those in which the axions appear, the N-flation mass
spectrum actually arises in this very simple way.

In the special case that the distribution of the $R_{Ai}$ is
Gaussian, $M$ is actually a real Wishart matrix \wishart, about
which a great deal is known.  In this paper we do {\it{not}}
assume that the distribution is Gaussian, as there is no clear
physical motivation for this beyond the universal behavior of
random matrices \bai\ in the $N \to \infty$ limit.  However, in
the event that exact results are desired for a toy model, the
Wishart ensemble could prove to be quite tractable.

\newsec{Monte Carlo Analysis of Random Mass Matrices}

In the preceding section we made a few assumptions about the
statistical properties of the K\"ahler-covariant derivatives
$D_{A}C_{i}$ and about the K\"ahler metric itself.  Moreover, we
argued that the essential structure of the mass matrix
\massmatrix\ is captured by \triv.  These simplifications led to a
universal result for the mass spectrum, the Mar\v{c}enko-Pastur
law \mpis.

In this section we use Monte Carlo methods to provide further
evidence for the conclusions of \S4.  In particular, we show that
the results are robust against relaxation of our assumptions.  We
construct a collection of mass matrices according to the full rule
\massmatrix, using a variety of input distributions for
$D_{A}C_{i},K^{AB}$.  The result is indistinguishable from \mpis,
providing a strong confirmation our analytic result.

To generate a random mass matrix $M_{ij}$, we choose the entries
of the component matrices $K^{AB}$, $C_{i}C_{j}$, and $D_{A}C_{i}$
from appropriate ensembles $\Omega_{K},\Omega_{C},\Omega_{H}$.
This procedure is clearly superior to postulating the distribution
$\Omega_{M}$ of the entries of $M_{ij}$ directly, because the
latter procedure produces no new information.  In contrast,
assembling the component matrices according to \massmatrix\ may be
expected to yield results that are substantially independent of
the (largely unknown) details of
$\Omega_{K},\Omega_{C},\Omega_{H}$, but that do depend on the
known structure of the matrix products in \massmatrix. We will see
that this expectation is justified.

\subsec{Properties of the Component Matrices}

Although the K\"ahler metric is determined by the intersection
numbers $C_{ijk}$ of the Calabi-Yau, the scarcity of examples
where all intersection numbers are known makes this an impractical
route to an understanding of the properties of a typical $K^{ij}$.
For this reason, our examples of ``K\"ahler metrics" were simply
random, symmetric, positive-definite matrices.

The derivatives $D_{A}C_{i}$ split into two classes.  For $A=a$,
{\it{i.e.}} when the derivative is with respect to a complex
structure modulus, we have \eqn\dcompl{ D_{a}C_{i} =
\partial_{a}K C_{i} + \partial_{a}C_{i}}
and we recall that $C_{i}$ may depend on the complex structure
sector through the prefactor $A$, which is a fluctuation
determinant in the case that the nonperturbative superpotential
comes from Euclidean D3-branes.  As we have stressed, the form of
this dependence for a general threefold is not known.

For $A=i$, {\it{i.e.}} in the K\"ahler sector, we have instead
\eqn\dkahl{D_{i}C_{j} = \partial_{i} K C_{j} - 2\pi C_{j}
\delta_{ij}} so that the average behavior of the diagonal elements
is distinguishable from that of the off-diagonal elements.

Observing that each of the three classes of individual terms
$\partial_{a}C_{i}$, $\partial_{A}K$, $C_{i}$ can be positive or
negative, we constructed samples of $D_{A}C_{i}$ by drawing these
terms from distributions that ranged over both positive and
negative numbers of order one.   We found that for a wide range of
properties of these distributions, the mean value $\mu$ of the
elements of the resulting matrices $D_{A}C_{i}$ is much smaller
than the standard deviation $\sigma$.  This is consistent with our
analytic demonstration, in \S4.1, that the mean value of these
elements is necessarily small compared to their standard
deviation.

\subsec{Results of Monte Carlo Simulation}

We generated thousands of examples of mass matrices using the
rules outlined in the previous section.  In particular, we varied
the distributions $\Omega_{K},\Omega_{C},\Omega_{H}$. For example,
in the case of $\Omega_{H}$ we considered normal distributions
with various values of the mean; uniform distributions; log-normal
distributions; chi and chi-square distributions; and many other
distributions assembled from these via matrix products.

The results of this simulation are as follows.  The distribution
of matrix elements is bimodal, with the off-diagonal entries
normally distributed around zero, and the diagonal entries
normally distributed around a positive mean.  As may be expected
from the central limit theorem, the simulated distributions are
extremely close to normal distributions for a wide range of input
parameters.

The positive mean of the diagonal terms is large compared to the
expected magnitude of the off-diagonal terms.  Specifically, we
find that the typical magnitude of off-diagonal elements is
smaller than the typical magnitude of diagonal elements by a
factor \eqn\sis{ \varepsilon = {1\over{\sqrt{N+P}}}}  The
numerical coefficient is of order one. This suppression is of
limited relevance for our primary goal of understanding the mass
spectrum, but it could be useful in other contexts, such as an
analysis of the full axion potential \manycos.  For this reason,
we provide an analytic derivation of this suppression factor in
Appendix A.

Finally, and most importantly, the eigenvalues are indeed
distributed according to the Mar\v{c}enko-Pastur law, as shown in
Fig. 2.  The shapes of the empirical eigenvalue distributions
agree extremely well with those predicted by \mpis, without any
need for tuning or fitting of parameters.

The spectra shown are for the simplest case, that of a Gaussian
distribution with mean zero.  In each of the other cases we
tested, the spectra are identical to those shown, except for the
presence of a {\it{single}} eigenvalue $N$ times larger than the
others.  The presence of this single large eigenvalue is familiar
from random matrix theory \bai.  In physical terms, the associated
field will rapidly move to its minimum and decouple from the
inflationary dynamics.  Because a spectrum with an extreme outlier
is not easily interpreted visually, we do not show an example.

\fullfig{Empirical eigenvalue spectra of 100 random $300 \times
300$ matrices constructed according to \triv, with $\beta=1/10$,
$1/3$, $9/10$. The horizontal axis is the mass-squared and the
vertical axis is the relative probability. The overlaid curves
show the Mar\v{c}enko-Pastur law in each case.
}{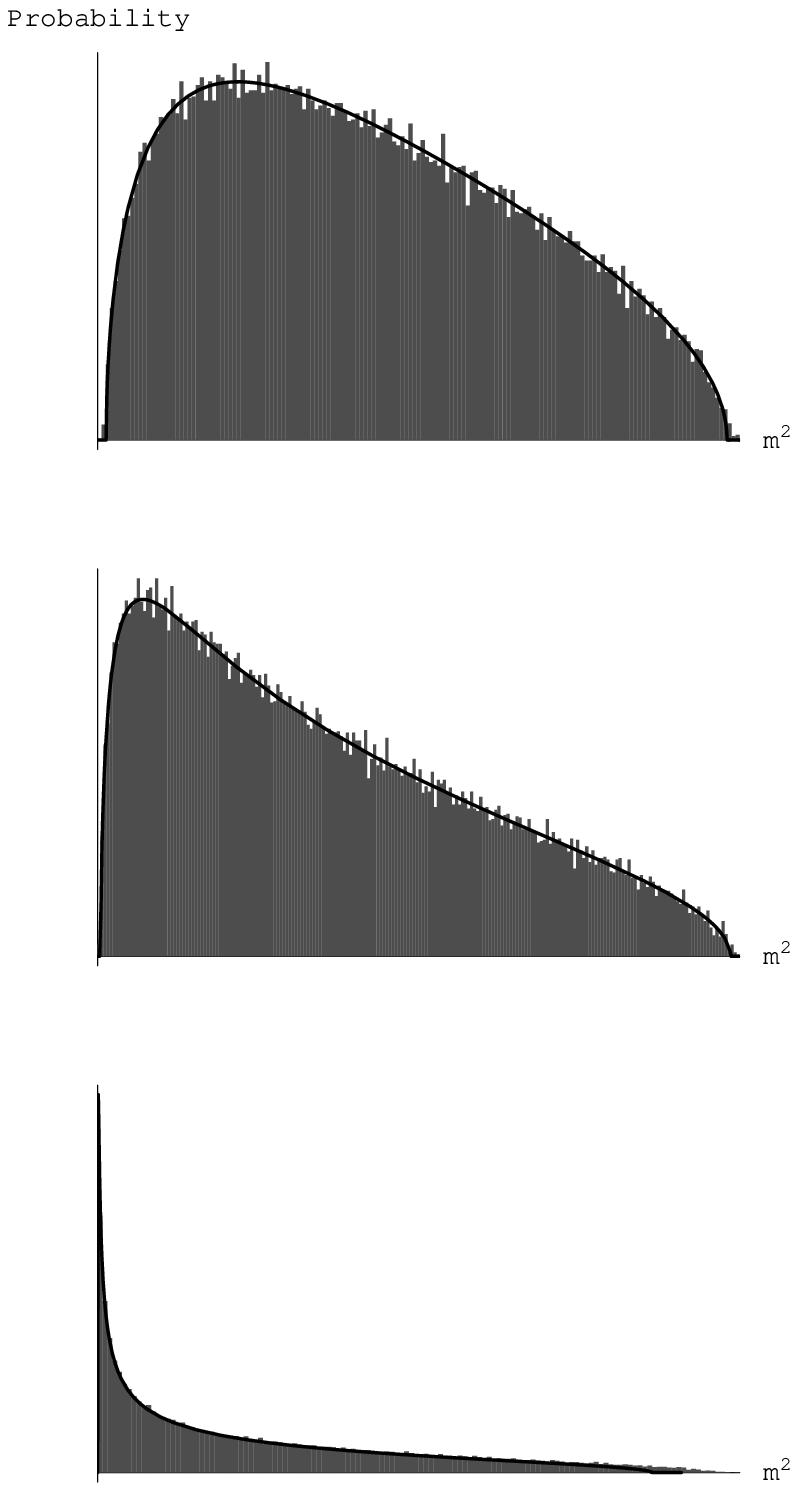}{4truein}

\def\mpl{M_{P}}

\def\Nfold{{\cal N}}

\def\PR{P_{\cal R}}

\newsec{N-flation Dynamics and Initial Conditions}

Having derived the mass spectrum for the fields contributing to
N-flation, we are now in a position to evaluate the cosmological
evolution associated with this system.

\subsec{Inflationary Dynamics}

As usual, the dynamics are described by the general set of
equations,
\eqn\Hsqrd{ H^2 =  {1 \over 3 \mpl^2 } \left[{1 \over 2} \sum_i
\dot{\phi_{i}}^2 + V(\phi) \right]  \, }
\eqn\Hdot{ \dot{H} = -{1 \over 2  \mpl^2 } \sum_i \dot{\phi_i}^2
\,  }
\eqn\phiddot{ \ddot{\phi}_i + 3 H \dot{\phi_{i}} +  {\partial V
\over
\partial \phi_i }  =0 \,  }
We will make frequent use of the fact that our potential $V$ is written as a sum over terms of the form $m^2_i \phi^2_i /2$.

During slow roll, one can find a general result for the field
evolution  \kaloperliddle. Dropping the acceleration terms from
the fields' equations of motion, and dividing the $\phi_i$
equation by the $\phi_j$ equation, we see
\eqn\phisr{
{\dot{\phi}_i \over \dot{\phi}_j}  =  {m_i^2\over m_j^2}  {\phi_i\over\phi_j}}
\eqn\phisrmass{ \Rightarrow    {\phi_i(t)\over\phi_i(t_0)} =
\left(   {\phi_j(t)\over\phi_j(t_0)} \right)^{m_i^2/m_j^2} \, }
This is not an attractor: if we rescale the initial value of one
of the fields by multiplying $\phi_i(t_0)$ by a constant,  this
rescaling changes the  field value at all subsequent times, for as
long as the slow roll approximation remains valid.

It is instructive to specialize to the Mar\v{c}enko-Pastur
spectrum, where the lightest field $\phi_{1}$ has $m_{1}^2 \approx
a$ \ais, the average mass-squared is $\sigma^2$, and the most
massive field $\phi_{N}$ has $m_{N}^2 \approx b$ \bis.  Let us use
the progress of $\phi_{N}$ as a clock, via \eqn\tauclock{ \tau(t)
\equiv {\phi_N(t)\over\phi_N(t_0)} } As inflation proceeds, $\tau$
decreases.  Trivially rewriting \phisrmass, we have
\eqn\applymass{{\phi_i(t)= \phi_{i}(0) \tau(t)^{Q_{i}}} \, } where
$Q_{i} = {m_{i}^2\over{b}}$ are in the range \eqn\vrange{
{a\over{b}} = \left({1-\sqrt{\beta}}\over{1+\sqrt{\beta}}\right)^2
\le Q \le 1} (In the preferred case $\beta \approx {1\over{2}}$,
we have ${a\over{b}} \approx {1\over{34}}$.)  Even if the fields
begin near the same point in field space, they quickly spread out:
the large range in the exponents $Q_{i}$ ensures that the fields
move at very different rates.  We illustrate this spreading in
Fig. 3.

If instead each field had a $\lambda_i \phi_i^4$ potential, the
solution analogous to \phisrmass\ would be
\eqn\phisrquart{
 {1\over\phi_i(t)^2} -  {1\over\phi_i(t_0)^2} =  {\lambda_i\over\lambda_j} \left(  {1\over\phi_j(t)^2} -     {1\over\phi_j(t_0)^2} \right) \,
}
We observe that the initial conditions become increasingly
irrelevant at late times, demonstrating the existence of an
attractor in the $\lambda \phi^4$ case.  This is a desirable
feature in a model of inflation, but because realizing assisted
$\lambda \phi^4$ inflation in string theory (or in any
well-motivated particle physics setup) is probably very difficult,
we will say no more about this case.

\fig{We plot the evolution of $\phi_i$ as a function of time for
$N=201$, $\beta=1/2$, and $i =1,21,\cdots,201$ with $\bar{m} = 10^{-6} \mpl$, and time measured in units where   $\bar{m} =1$.
}{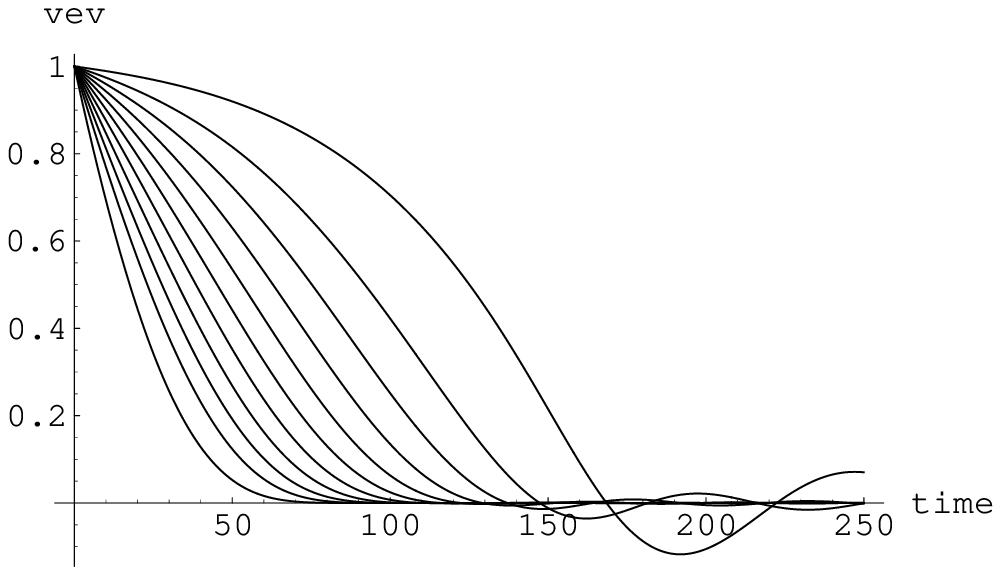}{5truein} The absence of an attractor solution for
N-flation significantly complicates the inflationary dynamics. If
there were a late-time attractor, and inflation lasted long enough
to eliminate the residual effects of the initial conditions, we
could assume that the system followed the attractor solution
during the last 60 e-folds of inflation, greatly simplifying the
treatment of the perturbations. By contrast, we will see in the
next section that the N-flation spectrum is a function of the
values of the fields measured at some fiducial time, since the
initial conditions never become irrelevant.

Finally, notice that in single-field inflation, the breakdown of
slow roll coincides with the end of inflation, and to a first
approximation the two events can be regarded as equivalent.  In
N-flation, however, slow roll can fail well before the end of
inflation for fields that are particularly massive or that begin
close to the origin.

\subsec{Initial Conditions}

We will not give a full treatment of the initial\foot{Throughout
this paper, we use ``initial'' to refer to the situation 60
e-folds before the end of inflation.  This is not necessarily the
start of cosmic history.} conditions in this paper. However, a few
observations will be essential.

The data of an N-flation model in the four-dimensional effective
theory is a spectrum of $N$ axion masses.  Given any such system,
one would like to know whether the model can produce, with some
choice of initial conditions, ${\cal{N}} \gtrsim 60$ e-folds of
inflation.  For this purpose it is useful to choose initial
conditions that maximize the number of e-folds while remaining
consistent with all control assumptions, especially the
requirement that the {\it{largest}} vev is at most of order
$M_{p}$.  The amount of inflation is maximized by placing all of
the fields as far from the origin as this condition will permit;
spreading the initial vevs inevitably moves some fields closer to
the origin and thus diminishes the number of e-folds.  N-flation
systems do not readily produce a vast number of e-folds, so only a
tiny fraction of N-flation models can generate enough inflation
from a highly ``inefficient'' initial configuration with disparate
vevs. For this reason, we will focus primarily on initial
conditions in which the fields are spread by at most an order of
magnitude.

Consider any distribution $\{\phi_{i}\}$ of initial field values,
and compute the correlation \eqn\simcor{ \xi \equiv {\langle m^2
\phi^2 \rangle\over{\langle m^2 \rangle\langle \phi^2\rangle}}
\equiv {N \sum_{i} m_{i}^2 \phi_{i}^2\over{{\sum_{j}
m_{j}^2\sum_{k} \phi_{k}^2}}}} where we have defined
\eqn\defdoubmom{ \langle m^{p} \phi^{q} \rangle \equiv {1\over{N}}
\sum_{i} m_{i}^{p} \phi_{i}^{q}}
When $\xi>1$ ($\xi<1$), we say that the vevs are directly
(inversely) correlated with the masses.

We may also parametrize the correlation between masses and vevs as
\eqn\otherpar{ \phi_{i} = \upsilon_{i} m_{i}^{\alpha}} where the
variation in prefactors $\upsilon$ is presumed to be less
important than the dependence on the mass.  Let us try to relate
$\alpha$ to $\xi$. We begin by observing that
%
for reasonable values of $N$, $\langle m^{2k}
\rangle$ is very well approximated as a moment of the
Mar\v{c}enko-Pastur distribution,
\eqn\defmpm{ \langle m^{2k}\rangle \approx m_{(2k)} \equiv  \int
m^{2k} p(m^{2})  }
where the approximation omits terms of order $1\over{N}$ \edel,
which we can safely neglect in the following.  The moments of
\mpis\ are in turn given by a simple analytic formula \hungar,
\eqn\mpmomentsare{ m_{(2k)} = \sigma^{2k} \sum_{i=1}^{k}
{1\over{k}}{k\choose{i}}{k\choose{i-1}}\beta^{i-1} \, }
We will also need to know $\langle m^{-2}\rangle$ and $\langle
m^{-4}\rangle$.  These are not moments of the distribution, but we
can compute them directly from the Mar\v{c}enko-Pastur law,
finding
\eqn\negtwomoment{ \langle m^{-2} \rangle = {1 \over \sigma^2} {1
\over (1-\beta)} \, }
and \eqn\negmoment{
 \langle m^{-4} \rangle = {1 \over \sigma^4} {1 \over (1-\beta)^3} \, }
Equipped with these results, we can compute $\xi(\alpha)$ in terms
of $\beta$ for the cases of interest.

As a consequence of \phisrmass, the fields rapidly spread apart
(except when the masses are degenerate), with the most massive
fields relaxing most quickly.   This means that the lighter fields
soon have larger vevs,
on average, than the heavier fields.  Thus, positive correlations
$\xi \gtrsim 1$ are transient, and inverse correlations $\xi < 1$
typically develop. This situation is quite plausible dynamically,
but, within the context of controllable N-flation models, it
creates a certain tension. Initial conditions with $\xi \gtrsim 1$
are the most efficient at producing many e-folds within the region
of control,\foot{It might be that inflation occurs in nature in a
way that is not theoretically computable; for obvious reasons we
do not consider this case.} but because such positive correlations
are typically transient, they may not be the most generic initial
conditions. We will work not with the initial conditions that are
most plausible {\it{a priori}}, but with those that are most
plausible given the requirement of ${\cal{N}} \gtrsim 60$ e-folds.

These considerations lead us to focus on initial conditions with
$\alpha \approx 0 \iff \xi \approx 1$, because anything else will
typically give inadequate inflation.  We will frequently refer to
this as an ``uncorrelated'' initial condition, because the vevs
are not correlated to the masses.  The equal-field condition
$\phi_{i}=\bar{\phi}$ is an important special case of uncorrelated
initial conditions.  Another instructive configuration, albeit an
impractical one for realistic models, is $\alpha=-1$, where each
axion makes a roughly equal contribution to the initial energy.


We have numerically evaluated the full equations of motion for a
wide variety of parameter values and initial conditions.  In Fig.
4 and Fig. 5 we show the results of the evolution for two specific
choices of initial conditions: the ``equal-field'' condition
$\alpha=0$, $\upsilon_{i} \approx \bar\upsilon$, and the
``equal-energy'' condition $\alpha=-1$, $\upsilon_{i} \approx
\bar\upsilon$.

\fullfig{N-flation dynamics with $N=200$, $\beta={1\over{2}}$, and
$\bar{m}  = 10^{-6} \mpl$, and equal-field initial conditions.
Time is measured in units where $\bar{m}=1$. The fields are
arranged along the axis labeled $i$, in order of increasing mass.
The top left plot shows the field evolution. The top right plot
displays the relative contribution of each field to the overall
energy density, normalized by the field that is currently making
the dominant contribution.  The bottom left plot shows the ratio
of the numerical solution to the slow roll solution, \phisrmass, computed in terms of the vev of the lightest field.
The bottom
right plot shows the evolution of the scale factor as a function
of time. }{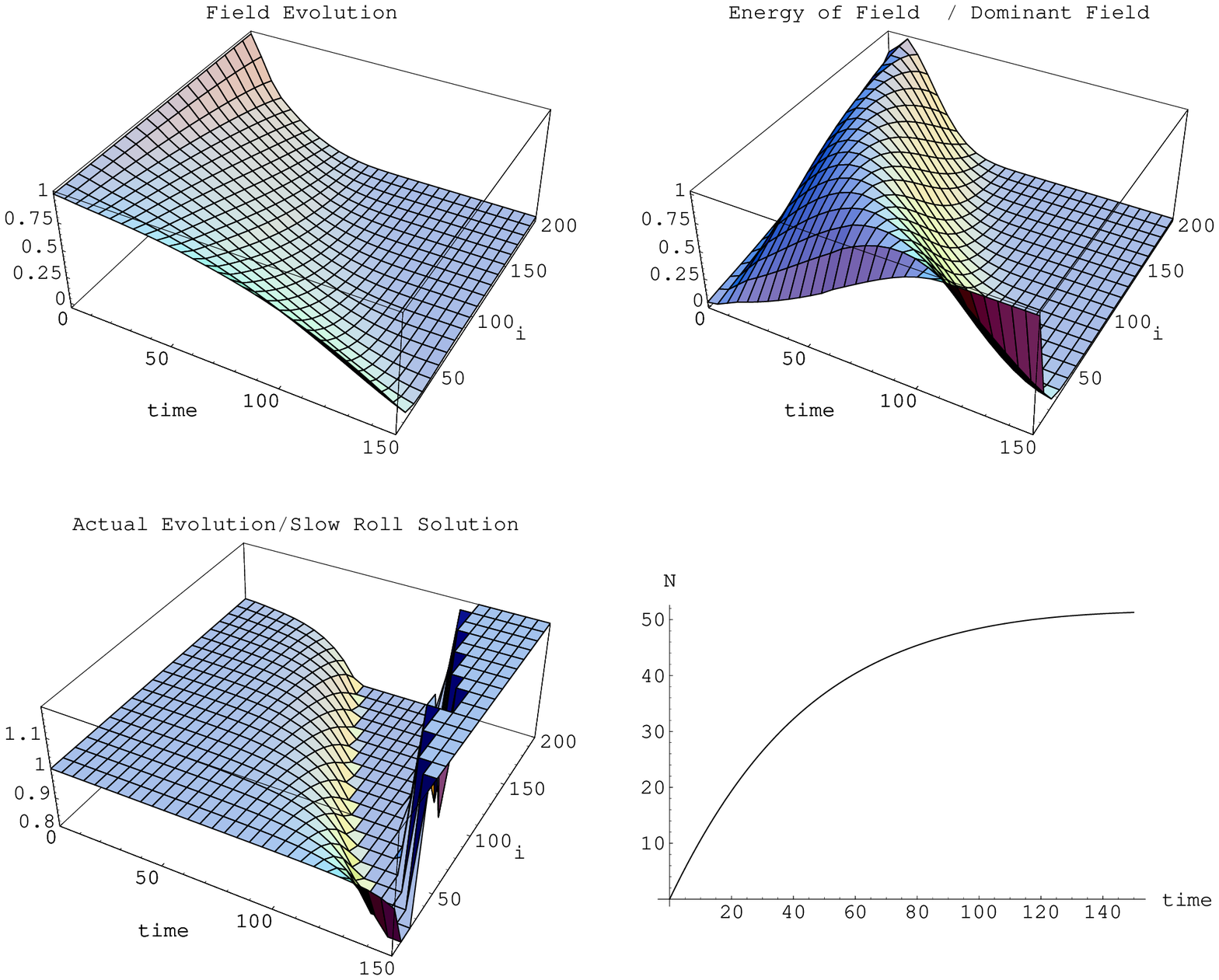}{6.3truein}
\fullfig{N-flation dynamics with $N=200$, $\beta={1\over{2}}$, and
$\bar{m}  = 10^{-6} \mpl$, and equal-energy ($\alpha=-1$) initial
conditions. The plots follow the layout of Figure 4.  This is an
illustration, not a controllable model: in order for the total
number of e-folds to match Figure 4, the initial vevs of the light
fields have been taken to exceed $\mpl$.  This is a concrete
example of the general observation that for a given mass
distribution, initial conditions with $\alpha< 0$ require larger
maximum vevs in order to produce enough
inflation.}{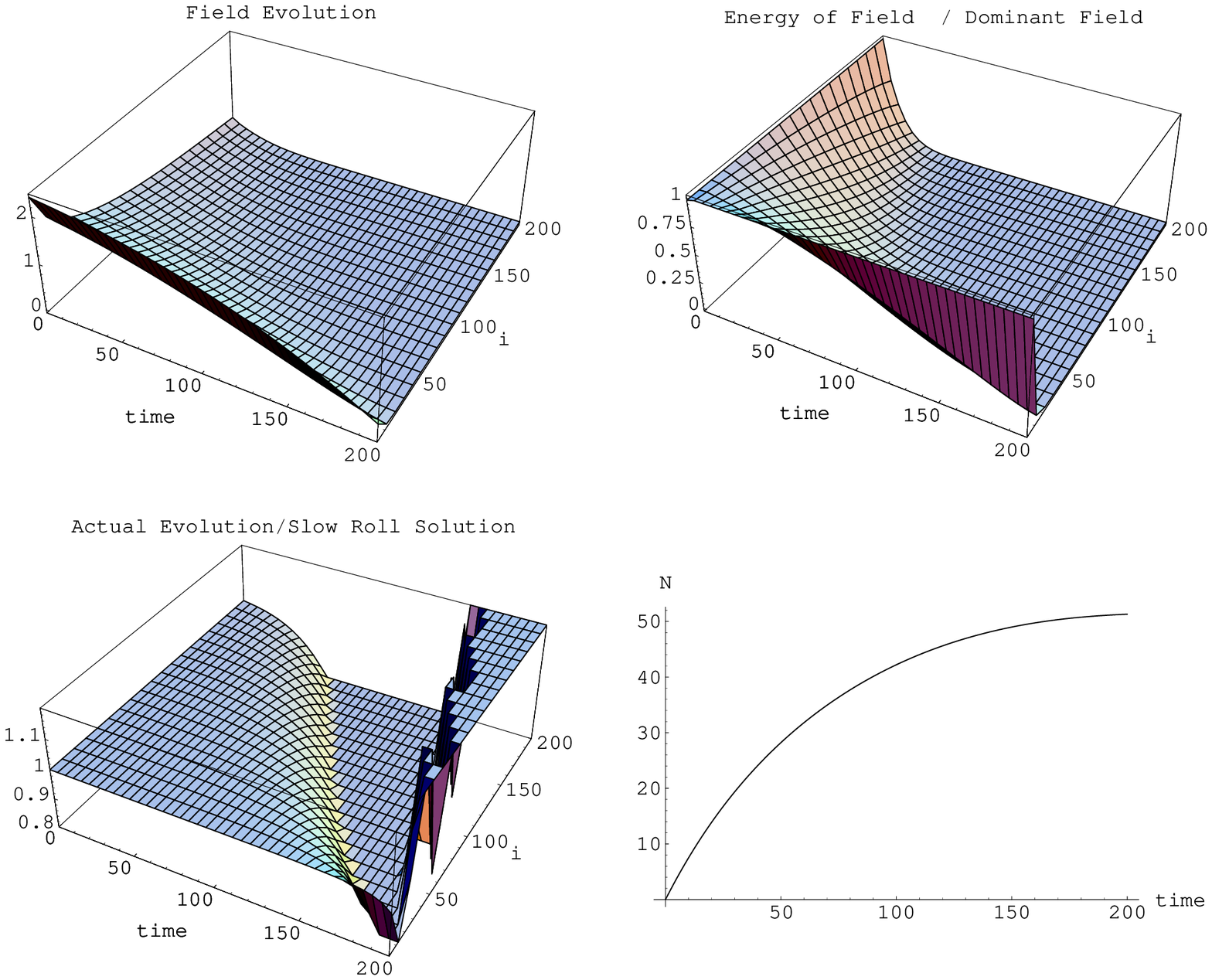}{6.3truein}

\subsec{Number of E-folds}

We will now study how the amount of inflation depends on $\beta$.
Armed with the general result \phisrmass\ for the field evolution,
we can write the potential $V(t)$ as an appropriately weighted
integral over the Mar\v{c}enko-Pastur mass distribution, since
\eqn\vtimeA{
 V(\phi_N(t)) = {1 \over 2 } \sum_i m_i^2 \phi_{i}^2(t_{0}) \tau(t) ^{2Q_{i}} \,
}
This is approximated by
\eqn\vtimeB{
 V(\tau(t)) = {\bar{\phi}^2 \over 2 } \int_a^b \lambda x^\lambda p(\lambda)
}
where \eqn\xdefA{ x = \tau(t)^{2/b}} and we have assumed that
$\phi_i(t_0) = \bar{\phi}$ for all $i$. However, this integral
cannot be done in closed form, so we need to make a further
approximation in order to compute the number of e-folds of
inflation we can expect for a given initial value $\bar{\phi}$. In
practice, we are particularly interested in moderate values of
$\beta$  -- as $\beta$ approaches unity, the mass distribution
develops a very long tail, but this situation is disfavored
physically. Consequently, we can get a {\it{rough}} estimate of
the number of e-folds by assuming that the $N$ fields have equal
masses $\bar{m}^2 = \sigma^2$ and possibly different initial vevs
$\phi_{i}$.
In this case, one can define $\Phi^2 \equiv N\langle \phi^2
\rangle$ and write down an equivalent single-field model \nf\ with
potential
\eqn\vsingle{ V(\Phi) = {1\over 2 } \bar{m}^2 \Phi^2 }
The power of assisted inflation is that each of the $\phi_i$ is
related to $\Phi$ by $\phi_i \approx \Phi /\sqrt{N}$, and while
the collective field $\Phi$ must exceed $\mpl$ by at least an
order of magnitude to obtain a workable period of inflation, the
vevs of the $\phi_i$ can all be sub-Planckian if $N$ is large
enough.

The total number of e-folds in this simple model is
\eqn\efoldssingle{ {\Nfold(\Phi)} = {1 \over  \mpl^2 } \left[ {
\Phi^2 \over 4}  -{\mpl^2 \over 2} \right] \approx {\Phi^2 \over 4
\mpl^2}  \, }
This turns out to be an excellent approximation to the number of
e-folds produced in the more general case of a spectrum of masses.
The quality of this approximation is demonstrated in Figure 6; we
conclude that it is safe to ignore $\beta$ when estimating $\cal
N$, unless $\beta$ is very close to unity, and provided that we
take initial conditions with approximately equal fields.

\fig{Numerical results for the number of e-folds $\cal N$ as a
function of $\beta$, with $N = 300$ and initial field values
$\phi_i = \mpl$. The expected number of e-folds from the
approximation in \efoldssingle\ is 75, so when $\beta$ is not very
close to unity, the approximation is
excellent.}{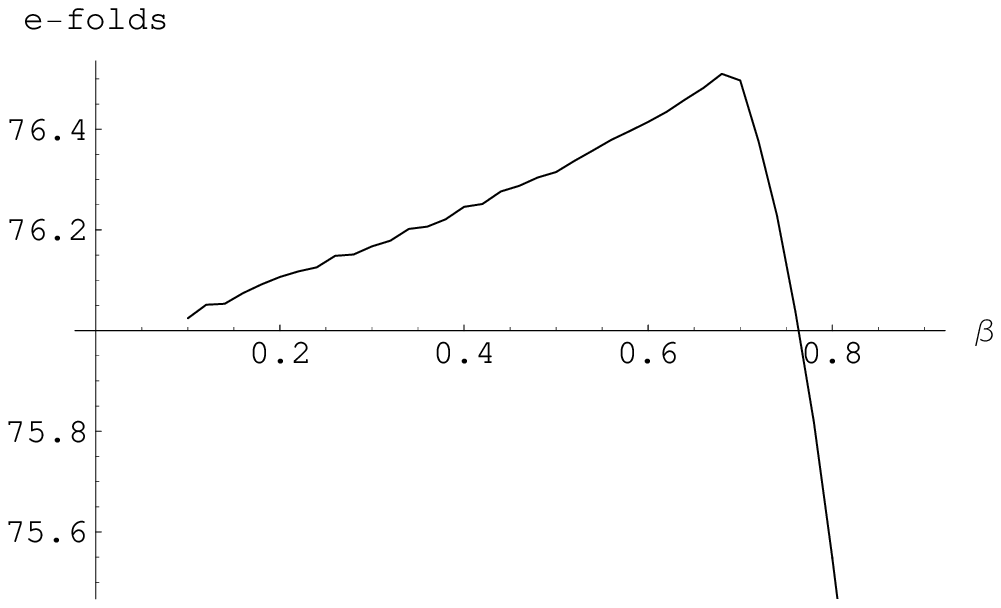}{4truein}

The number of e-folds also depends on $\alpha$.  As we are
physically constrained to consider only sub-Planckian vevs for the
$\phi_i$ we can compute the maximum number of e-folds by
saturating this bound for all the fields. Then any physically
realistic initial configuration will have $N$ vevs which are less
than or equal to their values in this idealized scenario, and the
amount of inflation produced when some vevs are initially reduced
from their maximal values will necessarily be lower.  We can
therefore find an upper bound on $\Nfold$ for a given mass
distribution by assuming that $\alpha =0$ and $\phi_i \approx
\mpl$.

Note that the mass scale $\bar{m}=\sigma$ has fallen out of
\efoldssingle.  In the end, the mass is fixed not by the number of
e-folds, but by the amplitude of the perturbation spectrum, which
we now discuss.

\newsec{Generation of Adiabatic Perturbations}

There is a very general result, derived in \Sasaki, for the
spectrum of curvature perturbations generated by multi-field
inflation,
\eqn\powerGeneral{ P_{\cal R} = \left({H \over 2\pi} \right)^2
{\partial \Nfold \over \partial \phi_i} {\partial \Nfold  \over
\partial \phi_j} \delta_{ij} \, }
where $\Nfold$ counts the number of e-folds.  We will follow the
treatment in \Lyth, which specializes \powerGeneral\ to the case
of uncoupled fields with polynomial potentials, \eqn\lythv{
V(\phi) = \sum_{i} V_{i}(\phi_{i}) } In the case of N-flation,
$V_{i} = {1\over{2}}m_{i}^2\phi_{i}^2$.

The spectrum is given by \Lyth\
\eqn\powerV{ \PR^{1/2} = \sqrt{{ V \over{12 \pi^2 \mpl^6}} \sum_i
\left({V_{i} \over V_{,i}} \right)^2} \,  }
We have chosen to work with $\PR^{1/2}$, since this is
proportional to the amplitude of the primordial density
perturbations, and to the amplitude of the temperature
perturbations in the microwave background.   We are implicitly ignoring
entropy (isocurvature) fluctuations in this initial analysis,
although these are likely to be subdominant for uncoupled fields
with very similar decay channels.

For our specific potential, the spectrum becomes
\eqn\powerNflation{ \PR^{1/2} =   {1 \over \mpl^3 } \sqrt{{1\over
12 \pi^2} \left(  \sum_i{  {1 \over 2} m_i^2
\phi_i^2}\right){\sum_i \left( {\phi_{i} \over 2} \right)^2}}}

Combining these results with \defdoubmom, we find \eqn\PRgen{
 \PR^{1/2} =   {N \over  \mpl^3 } \sqrt{{\langle m^2 \phi^2 \rangle \langle \phi^2
\rangle} \over 96 \pi^2}\, }

We stress that \defdoubmom\ is only a convenient notation, and
does {\it{not}} imply that an averaging has taken place.  The
expression \PRgen\ is exactly equal to \powerNflation, for any
collection of masses and field vevs.  Of course, our purpose in
writing \PRgen\ is to make use of statistical information about
the masses and initial conditions. To accomplish this, we will now
use the moments \mpmomentsare\ of the Mar\v{c}enko-Pastur
distribution, in combination with various possibilities for the
initial conditions.

In the case $\xi=1 \iff \alpha=0$ that the initial field values
are uncorrelated with the masses, which includes the important
special case of equal initial vevs, we have \eqn\nocorr{ \langle
m^{p} \phi^{q} \rangle = \langle m^{p} \rangle \langle \phi^{q}
\rangle} so that \eqn\PRsimp{ \PR^{1/2} = {{{N \langle \phi^{2}
\rangle}\sqrt{\langle m^{2} \rangle}}\over{\pi\sqrt{96}M_{p}^3}}}
If the masses are taken to be identical, this reduces to the
result of an effective single-field model, \eqn\simpsing{
\PR^{1/2} = {{\Phi^{2}m_{\Phi}}\over{\pi\sqrt{96}M_{p}^3}} =
{V(\Phi)\over{m_{\Phi}\pi\sqrt{24}M_{p}^3}}} once we identify
$\Phi^2= N\langle\phi^2\rangle$. This is the simplest possible
model of N-flation.

Considering instead the case $\alpha=-1$ in which each field makes
a roughly equal contribution to the energy, we find instead
\eqn\PRener{ \PR^{1/2} =
{V\over{\bar{m}\pi\sqrt{24}M_{p}^3}}{1\over{\sqrt{1-\beta}}}}
which is enhanced relative to the single-field result by the
modest factor $(1-\beta)^{-1/2} \sim \sqrt{2} $.

\fig{Scalar power spectra $\PR^{1/2}$, computed from
\powerNflation\ for the initial conditions used in Fig 4 (lower
line) and in Fig 5. The masses are identical in the two
cases.}{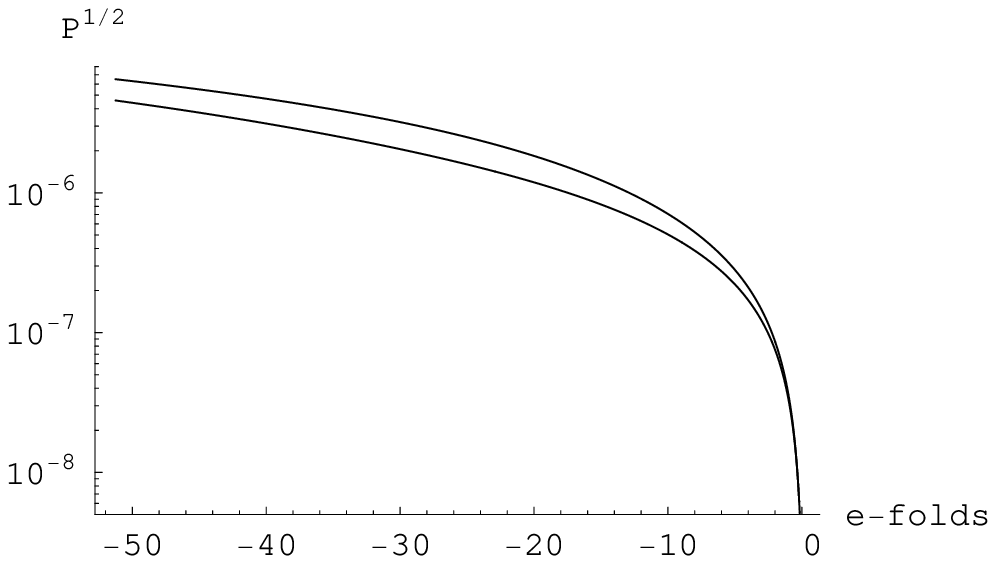}{4truein}

To get a clearer understanding of the connection between the
scale-dependence in the spectrum and the initial field
configuration, we now turn to the spectral index $n_s$, which is
defined as $n_s-1 \equiv d \ln \PR / d \ln{k}$, so that $\PR(k)
\sim k^{n_{s}-1}$.  For a single inflaton, one has as usual
\eqn\singlefieldn{ n_{s}-1= 2\eta - 6\epsilon} where
\eqn\etaepsilon{\epsilon = { \mpl^2 \over2 } \left( {V'\over V}
\right)^2 \qquad \eta =  { \mpl^2 } {V''\over V}} In the case of
single-field $m^2\phi^2$ inflation, this becomes \eqn\sfn{ n_{s}-1
= - {{8 M_{p}^2}\over{\phi^2}}} For multiple fields, $n_{s}$ obeys
the general formula \Sasaki\
\eqn\index{n_s-1 =  2{\dot{H} \over H^2} - 2 {{  \Nfold_{,i}
\left( {1 \over \mpl^2} {\dot{\phi}_i \dot{\phi}_j \over H^2 } -
\mpl^2 {V_{,ij} \over V} \right)   \Nfold_{,j}} \over {\delta_{ij}
\Nfold_{,i}\Nfold_{,j}  }  }\, .}
Using once again the specialized results of \Lyth\foot{This
simplification is only possible when all the fields are still
rolling slowly.} this reduces to \eqn\ngen{
 n_{s}-1 = -\mpl^2 \sum_i \left( {V_{,i}\over V}\right)^2 -
  { 2 \mpl^2 \over \sum_i \left(  {V_i \over V_{,i} }\right)^2 }
+   2 {\mpl^2 \over V}  {{\sum_i { V_{,ii}V_{i}^2 \over V_{,i}^2}}
\over \sum_j \left( {V_{j} \over V_{,j} }\right)^2 }}
which reverts to \sfn\ in the single-field limit. For the special
case of multiple quadratic potentials, the final term is
proportional to the second term, so
\eqn\ngenA{
 n_{s}-1 = -\mpl^2 \sum_i \left( {V_{,i}\over V}\right)^2 -
  {  \mpl^2 \over \sum_i \left(  {V_i \over V_{,i} }\right)^2 } \, .
} It follows that $n_{s}-1$ is never positive, and thus that the
spectrum is never blue, because we have written $n_{s}-1$ as a
combination of squares with a negative overall coefficient.

\fig{The spectral indices $n_s$ corresponding to the spectra
plotted in Fig 7, displayed as functions of $\log{a}$.  The case with equal initial vevs initially has a higher index.
 }{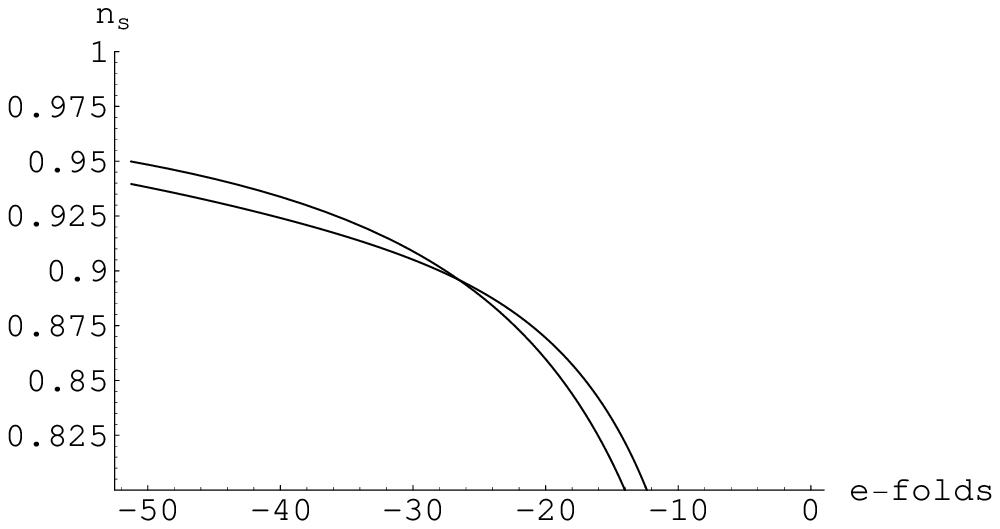}{4truein}

Using \defdoubmom\ we can rewrite \ngenA\ as \eqn\nsrewrite{
n_{s}-1 = -{4M_{p}^2\over{N}} \Bigl({1\over{\langle
\phi^2\rangle}}+{{\langle m^4\phi^2\rangle}\over{\langle
m^2\phi^2\rangle^2}} \Bigr)}
Again, we stress that \nsrewrite\
follows identically from \ngenA, rather than being a statisical
approximation. Results that do use statistical reasoning are
easily noticed because they contain factors of $\beta$, the
parameter that encapsulates the form of a typical mass
distribution.

For the case $\alpha=0$ of initial conditions that are
uncorrelated with the field masses, we use \nocorr\ to find
\eqn\nmpmom{ n_{s}-1 =
-{8M_{p}^2\over{N\langle{\phi}^2\rangle}}\Bigl(1+{\beta\over{2}}\Bigr)}
This coincides with \sfn\ only in the degenerate limit $\beta \to
0$.  Thus, for this important class of initial conditions, the
generalized N-flation spectrum is {\it{more red}} ({\it{i.e.}}
further from scale-invariant) than the single-field spectrum. This
is one of our main results.

In Fig. 9 we show the $\beta$-dependence of the spectrum \nmpmom\
in the case $\alpha=0$.  Because the result shown is the spectral
index more than 60 e-folds before the end of inflation, the values
shown are closer to scale-invariant than in
observationally-relevant cases.  Our purpose in showing the figure
is establishing the perfect agreement between \nmpmom\ and the
result of the  
numerical evaluation of \ngenA.

\fig{ The initial value of the spectral index as a function of
$\beta$, with all $N=300$ fields initially equal to $\mpl$. The
solid line is the theoretical curve \nmpmom, and the dots come
from numerical evaluations of \ngenA.}{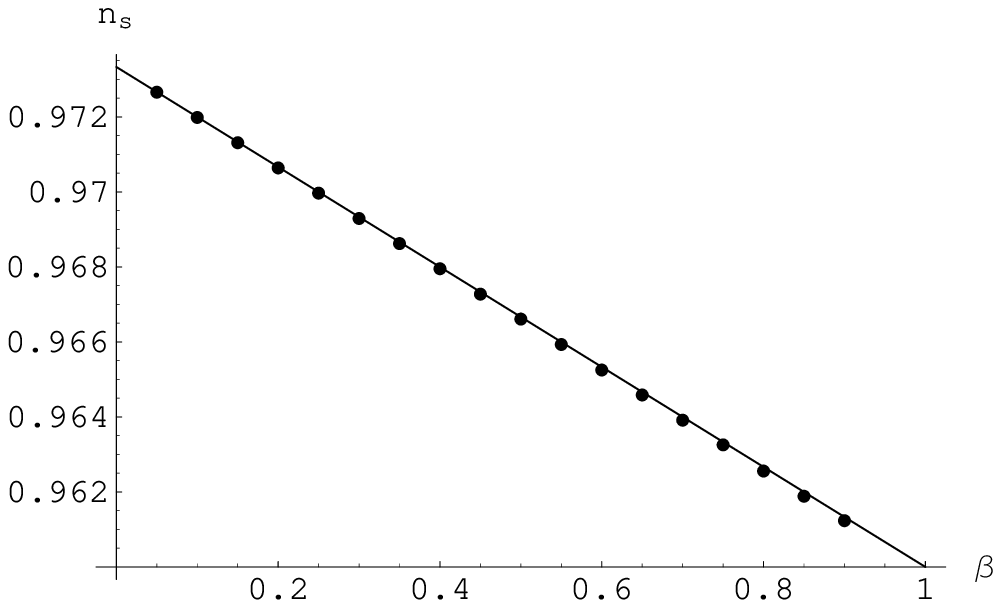}{4truein}

The quantitative relationship between $n_{s}$ and $\beta$ depends
on the initial conditions, and in particular on the correlation
$\alpha$ between the masses and the initial vevs.  As we stressed
in \S6.2, the displacements might become correlated inversely with
the masses, because the more massive fields relax more rapidly. It
would be very interesting to undertake a careful study of the
initial conditions for N-flation, and to extract $n_{s}$ as a
function of $\alpha$ and $\beta$.  However, this is a problem
beyond the scope of the present paper.

%
%
%
%
%
%
%

Finally, we can compute the running $ d n_{s} / d \ln{k}$ of the
scalar power spectrum, by applying the above methods to the
general formula given in \Lyth.  The leading contribution is
\eqn\running{ {{d n_{s}} \over { d \ln{k}}} =
-{16\over{N^2}}\Bigl( 2 { \langle m^4 \phi^2 \rangle^2 \over
\langle m^2 \phi^2 \rangle^4} - { \langle m^6 \phi^2 \rangle \over
\langle m^2 \phi^2 \rangle^3}+ {1\over{\langle \phi^2 \rangle^2}}
\Bigr)} For the uncorrelated initial condition $\alpha=0$, this
simplifies to \eqn\easyrun{ {{d n_{s}} \over { d \ln{k}}} = -{16
M_{p}^4\over{N^2 \langle \phi^2
\rangle^2}}\left(2+\beta+\beta^2\right)} This reproduces the
single-field result if we again identify $\Phi^2=
N\langle\phi^2\rangle$ and send $\beta \to 0$.  Evidently the
Mar\v{c}enko-Pastur spread in masses creates a modest increase in
the running, by a factor that is at most $2$ and is typically
around $11 \over 8$.

This analysis is far from being the last word in understanding the
inflationary phenomenology of N-flation. To fully understand the
spectrum arising from an arbitrary set  of initial conditions, it
will be necessary to go beyond first order in slow roll -- and
perhaps to numerically solve the evolution equations for the field
perturbations, since the heavy fields are not well described by
the slow-roll approximation during the last e-folds of inflation.
Moreover, because there is no attractor solution and the spectrum
is a function of the initial field configuration, a solid
prediction for the spectral index will require an understanding of
the expected distribution of initial values for the axion fields.
A further complication we do not consider is that the value of $H$
at the end of inflation depends both on $\beta$ and on the initial
values of the axion vevs.  In general,  increasing $\beta$ while
holding the vevs and the overall mass scale fixed will lower the
value of $H$ at the end of inflation.  This effect must be
incorporated in a full calculation of the perturbation spectrum,
as it  changes the number of e-folds before the end of inflation
at which a given comoving scale leaves the horizon, modifying the
predicted spectral amplitude for that mode.
Finally, it will be important to understand the possible role of
isocurvature fluctuations.

\newsec{Astrophysical Constraints on the N-flation Parameters}

The  simplest realization of N-flation is that in which $N$
uncoupled, canonical fields have identical masses.  As explained
in \nf, this model is equivalent to chaotic inflation with the
potential $m^2\phi^2$.  On the other hand, we have also seen that
a spread in the masses will modify the scale-dependence of the
spectrum.  We now turn to the observational constraints that can
be imposed on the parameters of N-flation. Since the spectrum
depends on the initial field distribution, we will focus on
qualitative constraints, rather than constructing detailed
exclusion diagrams.

First, we can normalize the mass scale by comparing $\PR$ to the
amplitude of the temperature fluctuations observed in the CMB
\PeirisFF:
\eqn\PRobs{ \PR =  \left({5 \over 3}\right)^2 {800\pi^2 \over
T^2_{ CMB}} A(k_0) =
 2.95 \times 10^{-9}  A(k_0)}
where $A(k_0)$ parametrizes the height of the spectrum at some
fiducial wavenumber $k_0$ and $T_{ CMB}$ is expressed in
micro-Kelvins.   Observationally, WMAP measures $A \approx  0.78$
\refs{\BennettBZ,\SpergelCB}.  For any given initial conditions
the mass $\bar{m} \equiv \sigma$ is determined by comparing
\PRgen\ and \PRobs; however, the precise result can change by a
factor of a few, depending on $\alpha$ and $\beta$. Taking $\Nfold
\approx 60$ and initial conditions that are not correlated with
the masses ($\alpha=0$), we find
\eqn\PRdata{ \sigma  \approx 1.5 \times 10^{-5} \mpl }

In addition to the scalar perturbations, we must also consider the
production of tensor perturbations during inflation.  These are
generated by quantum-mechanical fluctuations in the spacetime
background itself, and have amplitude proportional to $H \propto
\sqrt{V}$. One can seek these at CMB scales, through their contribution to the temperature anisotropy,
through their contribution to the polarization B-mode (see
{\it{e.g.}} \UngarelliQB\ for the likely sensitivity of future
experiments), or via a direct detection of a stochastic background
of gravitational waves with a mission such as BBO \BoyleSE.   Like
the single-field $m^2 \phi^2$ model it mimics, N-flation occurs at
a comparatively high scale, and thus yields a significant tensor
spectrum.  However, to break the degeneracy between N-flation and
single-field $m^2 \phi^2$ inflation, a very accurate measurement
of the tensor spectrum would be required.  The exact relation
between the tensor perturbations in the single-field model and in
N-flation will depend on the distribution of initial vevs, as well
as on $\beta$.  In practice, however, the amplitudes are certainly
of the same order in the two cases, though not necessarily
identical.  Moreover, as both N-flation and $m^2 \phi^2$ produce a
large tensor spectrum relative to most other inflationary models,
a sufficiently tight constraint on the tensor
 amplitude would rule
out both N-flation and single-field $m^2 \phi^2$ models.

The next constraint we explore is the lower bound on $N$, arising
from \efoldssingle\ and from the requirement that the individual
axion vevs $\phi_i$ never approach $\mpl$. Again taking
${\cal{N}}\sim60$, we find
\eqn\Nlimit{ {\phi_i^2 \over \mpl^2 } \sim {240 \over N  }}
at the onset of inflation.  Unless one can find reasonable string
theory constructions that allow values of $N$ as high as $10^4$,
the axion vevs in a model that produces adequate inflation will
have to begin uncomfortably close to the Planck scale, though
still below it.  Although this situation is not entirely
satisfying, it is still preferable to a single-field model in
which the initial vev is parametrically large compared to the
Planck scale, as it is improbable that models of the latter sort
can exist in any controlled framework in string theory.

In the previous section, we showed that the spectral index is
similar to that produced by single-field $m^2 \phi^2$ inflation,
but with somewhat more tilt at CMB scales (at least for those
initial field configurations we considered explicitly.)  These
changes in the scalar power spectrum can be noticeable in the
theoretically-favored case of the Mar\v{c}enko-Pastur distribution
with $\beta \sim {1\over 2}$, in that $n_s-1$ can change by around
25\%. Observing this difference unambiguously would
require a CMB / Large Scale Structure dataset of exquisite
precision, but is potentially within the reach of anticipated CMB missions.

In addition to curvature perturbations, multi-field inflationary
models can also generate isocurvature (entropy)
perturbations,\foot{See \ByrnesTH\ for an interesting treatment of
isocurvature in an assisted chaotic inflation model with $O(N)$
symmetry. This analysis corresponds to the $\beta \rightarrow 0$
limit in our terminology. With $\beta \approx 1/2$ the masses vary
significantly and the $O(N)$ symmetry is badly broken.
} as well as non-Gaussianities  \RigopoulosUS\ beyond those found
in the single-field case. In this initial analysis we have ignored
both of these effects, along with the possible impact of
higher-order corrections to the slow roll formalism.  Taking these
issues in reverse order, working at lowest order in slow roll
gives  us a very good account of the inflationary phenomenology of
the single-field $m^2 \phi^2$ model. In N-flation the spread in
the mass spectrum along with a generic, non-identical, set of
initial vevs for the axions means that some fields will cease to
be critically damped before the inflationary era comes to an end,
{\it{i.e.}} before the second derivative of the scale factor
$a(t)$ becomes non-positive.  If the initial vevs are similar and
$\beta$ is not very close to unity, this breakdown in slow roll
will likely only occur near the end of inflation, but in the
general case one needs to exercise caution when using any
slow-roll results. Further, understanding the non-Gaussianities
\RigopoulosUS\ and isocurvature modes \bart\ generated during
multi-field inflation requires a calculation that extends well
beyond the slow-roll approximation of the curvature perturbation,
but may provide further avenues for breaking the degeneracy
between N-flation and its single-field limit.  We intend to address this problem directly in future work.   
In practice, a
definitive understanding of the perturbation spectrum might
require evolving the $N$ fields' perturbation equations alongside
the background equations, as was done in \AdamsVC\ for the
single-field case.  This would not be prohibitively challenging:
the $N$-field simulations displayed in this paper take a few
seconds to complete, and adding the perturbations would roughly
double the number of relevant degrees of freedom.  Thus, the
generic numerical problem may be computationally tractable.

Finally, it is encouraging to note that astrophysical data is
rapidly reaching the point where it can put severe constraints on
large classes of inflationary models.  In particular, the recent
analysis by Seljak {\it{ et al.}} \SeljakXH\ uses a  large set of
astrophysical data sources to argue for tight constraints on the
primordial spectrum, which rule out many ``standard'' models of
inflation. In particular, $\lambda \phi^4$ is excluded at the
3$\sigma$ level, while the single-field $m^2 \phi^2$ model makes
predictions that differ from the central value at the 2$\sigma$
level.  Obviously these constraints on the primordial spectrum
will need to be confirmed (and will hopefully tighten) as further
datasets become available. As noted earlier, we have not fully
analyzed the inflationary predictions of N-flation in the general
$\beta \ne 0$ case, and we do not compute direct astrophysical
constraints on the parameter values.  However, the overall size of
the error ellipses in \SeljakXH\ suggests that observations may
soon put tight constraints on the N-flation parameter space (if
they do not do so already), providing further motivation for a
careful analysis of the cosmological predictions of this model.

\newsec{Possible Exceptions to our Results}

In this section we collect the conditions and assumptions that
were required in order to ensure the controllability of N-flation
generally and to derive the Mar\v{c}enko-Pastur spectrum.

First, in \S9.1, we review the constraints implied by controllable
perturbative and instanton expansions.  These requirements, which
were stated in \nf, have little to do with the inflationary
dynamics {\it{per se}}; instead, they should be thought of as
necessary conditions for a controllable trans-Planckian vev
created by N string axions.

Then, in \S9.2, we summarize the assumptions and constraints that
apply particularly to our random matrix ensemble of N-flation
models, with the intention of revealing the exceptional classes of
N-flation models that might avoid our conclusions.

\subsec{Conditions for a Trans-Planckian Displacement}

One of the main results of \nf\ is the construction of plausibly
trans-Planckian vevs $\Delta \Phi > M_{p}$ from multiple string
axions. As explained in detail in \nf, radiative corrections
diminish the effective size of $\Delta \Phi$ and reduce the
expected scaling \eqn\hopefor{ (\Delta \Phi)^{2} \propto N } to
\eqn\reallyget{ (\Delta \Phi)^{2} \propto {N\over{\chi(M)}} }
where we recall that in our notation, $P-1$ is the number of
complex structure moduli, and \eqn\defeuler{ \chi(M) = 2 N -
2(P-1) } The result is that $\Delta \Phi/M_{p}$ can be large, but
{\it{only}} when there are partial cancellations between $N$ and
$P$. This is a powerful reason to prefer models with $\beta \sim
{1\over{2}}$, where this cancellation is almost total. (An
unfortunate consequence of this numerical, rather than parametric,
success is that factors such as $2\pi$ are of considerable
importance in determining the output of a given model.)

Furthermore, although one might hope to construct trans-Planckian
vevs in a string vacuum with extremely weak coupling and extremely
large volume, this has not yet been achieved.  The problem is that
the axion decay constants diminish as the volume increases, so it
is not possible to work at arbitrarily large volume.

Let us therefore examine the conditions for control of the two
perturbative expansions of string theory: the string loop
expansion, controlled by $g_{s}$, and the sigma model expansion,
controlled by $\alpha'/R$, where $R$ is a typical length-scale in
the compact space.  Except when noted, we are here simply
summarizing the arguments of \nf.\foot{We thank J. Maldacena for
very instructive discussions of these arguments.}

The leading contribution to the Planck mass from sigma model
corrections is \eqn\delsig{{\delta_{\alpha'} M_{p}^2 \over{
M_{p}^2}} = {\chi(M)\zeta(3)\over{8\pi^3}}
{\alpha'^3\over{V_{6}}}} which depends on $N$ through $\chi(M)$.
Next, the dangerous corrections to $M_{p}$ from the string loop
expansion are those that have the leading N-scaling, namely
\eqn\delstrin{ {\delta_{g_{s}} M_{p}^2 \over{ M_{p}^2}} =
\sum_{q}\Bigl({{g_{s}^2 N}\over{16\pi^2 \gamma_{q}}}\Bigr)^{q}
\left({M_{kk}\over{M_{p}}}\right)^{2q}} In the final expression,
we have included the usual field theory loop factor of $16\pi^2$,
as well as a parameter $\gamma_{q}$ that measures the deviation of
the actual prefactor, in the string theory q-loop result, from
this naive expectation. (In \nf, the very conservative choice
$\gamma_{q} = {3\over{8\pi}}$ was made.  It would be worthwhile to
compute this factor directly for the one-loop case.) The ratio of
the Kaluza-Klein mass $M_{kk}$ to the Planck mass arises in
\delstrin\ for the following reason.  We are considering loop
corrections to the four-dimensional Planck mass that arise from
axions, or their moduli partners, circulating in loops.  These
fields have couplings of gravitational strength, and the diagrams
must be cut off at the compactification scale; combining these
results leads to \delstrin.\foot{In \nf, higher-loop terms were
conservatively taken to be proportional to a {\it{single}} power
of ${M_{kk}^2\over{M_{p}^2}}$. Even in this case, the requirements
of control can be met.}

The {\it{minimal}} requirements for perturbative control are
\eqn\minreq{ {\delta_{\alpha'} M_{p}^2 \over{ M_{p}^2}}< 1 }
\eqn\bminreq{ {\delta_{g_{s}} M_{p}^2 \over{ M_{p}^2}}< 1 } which
imply constraints on $g_{s}$ and on the compactification volume.
Among these is the important result that \minreq\ is most readily
satisfied when $\beta \sim {1\over{2}}$.

In addition, we must ensure that the single-instanton contribution
to the superpotential is a good approximation.  This follows from
\eqn\rhocon{ \rho_{i} > 1 \qquad i=1\ldots N} A skeptical reader
may be concerned that we have dropped multi-instanton terms in the
superpotential \wis\ but retained the products of two
single-instanton terms in the potential.  This is not
inconsistent: the leading contributions to the potential from a
multi-instanton term in the superpotential have the schematic
forms \eqn\leadw{ \Delta V \propto D_{a}W_{0}~e^{-2\pi(\tau_{i} +
\tau_{j})} + c.c. } where $a$ is a complex structure modulus, and
\eqn\oleadw{ \Delta V \propto W_{0}~e^{-2\pi(\tau_{i} + \tau_{j})}
+ c.c. } The factor $D_{a}W_{0}$ in the former would vanish \GKP\
if the background fluxes were imaginary self-dual.  In a vacuum
stabilized by nonperturbative effects, this factor will be
nonvanishing, but nonperturbatively small \nottwoone. Thus,
\leadw\ is negligible compared to the terms we have retained.
Next, \oleadw\ is highly suppressed because of the smallness of
$W_{0}$: to set an appropriate overall mass scale, we needed to
work with a flux configuration in which $W_{0}$ is very small \bp,
exactly as in \kklt.  Finally, notice that our random matrix
arguments could be extended even to cases in which
multi-instantons are relevant.

It was argued in \nf\ that the intersection of the constraints
\minreq,\bminreq,\rhocon\ is nonempty and includes configurations
with trans-Planckian displacements.  Even so, verifying this by
direct computation in an explicit example would be most valuable.

\subsec{Conditions for an N-flation Model with Mar\v{c}enko-Pastur
Spectrum}

For definiteness, we have focused our attention on KKLT vacua in
type IIB string theory.  However, scenarios other than that of
KKLT have some notable virtues for N-flation.  Type IIA moduli
stabilization with fluxes \twoa\ produces a hierarchy between the
scale of moduli stabilization and the scale at which
nonperturbative effects produce axion masses. This reduces the
challenge of arranging that the K\"ahler moduli $\rho_{i}$ are
massive enough to remain at their minima during inflation.
Heterotic string vacua have the advantage that the constraints of
perturbative control reviewed in \S9.1 can be less stringent for
NS axions than for RR axions, because the relation between the
compactification volume and the decay constants is different, by
factors of the string coupling, in the two cases. The axions
suitable for N-flation in known, stable type II vacua are RR
axions whose nonperturbative potentials come from D-brane
instantons, whereas in the heterotic string the relevant axions
come from the NS B-field and receive their potentials from
worldsheet instantons.

More importantly, for any vacuum in the above classes, the
{\it{only}} essential property that we provided as an input to the
supergravity potential \massis\ is that the axion potential arises
from F-terms that vanish when the axions are at their origin.  As
explained in \S2.3, in practice we require only the natural
assumption that any F-term energy present at the axions' minimum
is negligibly small compared to the energy scale of inflation. In
this sense, our results are extremely general: we expect that the
Mar\v{c}enko-Pastur spectrum of masses is characteristic of
N-axion potentials in most of the approximately F-flat vacua in
any supergravity theory.\foot{Although our primary motivation was
inflation, \mpis\ is truly a result about the N-axion potential,
whether or not this is relevant for inflation.}

Axion inflation around a vacuum that is not even approximately
F-flat could lead to a mass matrix whose form is different from
\triv. However, we expect that in such a case the mass spectrum
will be an appropriate mixture of the Wigner semicircle and
Mar\v{c}enko-Pastur distributions. Given a specific,
well-motivated model that is not F-flat, one could easily apply
our methods to determine $n_{s}$.

We approximated the entries \risnow\ as independent, random
variables. Strong correlations among these terms would violate
this assumption.  However, it was shown in \evencorrelated\ that
the Wigner law persists in ensembles with substantial correlations
among the matrix entries, and it is reasonable to anticipate a
similar result for the Mar\v{c}enko-Pastur law.  We are unaware of
any motivation for a correlation that is substantial enough to
invalidate \mpis. Of course, it would be extremely interesting to
find exceptions; the effect of known correlations could possibly
be analyzed through methods similar to those of \corr.  Moreover,
even in the presence of strong correlations and totally non-random
behavior, some basic results are available. Using only the fact
that $M$ \rotmassmatrix\ is positive-definite, and making no
assumptions whatsoever about the statistical properties of its
entries, it follows \mathmat\ that for any $i \neq j,$
\eqn\lspread{ m^2_{max}-m^2_{min} \ge 2|M_{ij}|} where the l.h.s.,
the difference between the largest and smallest eigenvalues of
$M,$ is known as the {\it{spread}} of $M$.  This implies a weak
but unavoidable lower bound on the difference between the largest
and smallest axion masses-squared.

We did not have to assume that the distributions are Gaussian or
otherwise simple, or that the mean $\mu$ is zero, because the
result of Mar\v{c}enko and Pastur can be shown \bai\ to follow
without these conditions. We also did not have to take the $N \to
\infty$ limit. The Mar\v{c}enko-Pastur law applies as $N \to
\infty,$ but is an excellent approximation for the values of $N
\gtrsim 200$ that were already required for N-flation to be
successful. Corrections to the moments \mpmomentsare\ are of order
$1\over{N}$ \edel\ and are too small to be relevant in our
results.

We did have to make a restrictive assumption (as was done in \nf\
for simplicity, though some statements made there do apply to
general initial conditions) about the initial conditions, namely
that the initial vevs obey $\phi_{i} \ll 1 ~~\forall i.$
The motion of $N$ axions with large vevs is governed by \manycos,
but we have left a study of the dynamics of such a system as an
interesting problem for the future.

We noted that, as a result of the renormalization of the Newton
constant explored in \nf\ and reviewed in \S9.1, models with
$\beta \sim {1\over{2}}$ are preferred. In such models the
spectrum is typically more red, for a range of initial conditions,
than in single-field $m^2\phi^2$ inflation. An obvious way to
avoid this constraint would be to demonstrate that models with
much smaller values of $\beta$ are in fact controllable and can
give enough inflation.

\newsec{Conclusions}

We began with the observation that substantial cross-couplings
between axions are a generic feature of N-axion models in string
theory. Moreover, complicated kinetic terms further connect the
various axions. Finally, as suggested in \nf, the masses of the
individual fields are not identical, in general, but are spread
over some range. Each of these effects can lead to nontrivial
departures from the simplest realization of assisted inflation:
the fields do not roll independently, and the dynamics is more
complicated than that of chaotic inflation with the potential $m^2
\phi^2$ \chaotic.

When the initial axion vevs are taken to be small in units of the
appropriate decay constants, the cosine potentials may be
expanded, and the leading term in the resulting potential is
quadratic.  In this setting, one may rotate and rescale fields to
produce canonical kinetic terms and a diagonal mass matrix.  The
result is a model specified by a spectrum of masses for $N$
uncoupled axions.

In this basis in which the axion kinetic terms are canonical and
the mass matrix is diagonal, the dynamics appears relatively
simple, though not quite as simple as in ordinary $m^2\phi^2$
inflation.  The most massive axions fall to their minima first,
diminishing the Hubble friction and releasing progressively
lighter axions to fall to their minima.  The Hubble parameter, and
hence the spectral index of the scalar perturbations, therefore
depends on the spacing in mass between adjacent axions. In any
given compactification, this spacing depends on a formidable array
of microscopic details. However, the statistics of the spread in
masses can be characterized much more simply.

We used random matrix techniques to compute\foot{The overall mass
scale $\bar{m}$ must be fixed by requiring the correct amplitude
of density perturbations, as in \nf.} the typical axion mass
spectrum, taking as input the structure of the supergravity
potential.  We found that the spectrum is given by a simple and
universal analytic formula \mpis, the Mar\v{c}enko-Pastur law \mp,
in the limit $N \to \infty$. For realistic values of $N$, the
agreement between this prediction and the results of Monte Carlo
simulations of the mass spectrum is excellent.

The shape of the Mar\v{c}enko-Pastur curve \mpis\ depends on a
single rational parameter $\beta$ \betis, which is determined by
the dimensions of the K\"ahler and complex structure moduli
spaces.  Our results and constraints are therefore phrased in
terms of $\beta$ alone. This is a rather dramatic simplification
from the naive expectation that a randomly chosen model of the
form \lgs\ would be characterized by ${\cal{O}}(N)$ independent
parameters. Even worse, in the full model \manycos\ in which the
initial displacements are so large that terms beyond quadratic in
the axions must be kept, at least ${\cal{O}}(N^2)$ parameters
would be required.  In this sense, random matrix models of
N-flation are surprisingly predictive.

We then studied the inflationary trajectory of a general N-flation
model.  Using the distribution \mpis\ to generate an ensemble of
mass matrices simulating an ensemble of N-flation models, we
numerically integrated the equations of motion for this system.
Next, we studied the adiabatic perturbations in a general
N-flation model.  We gave analytical expressions for the scalar
power spectrum \powerNflation\ and its tilt \nsrewrite\ and
running \running\ from which, given any spectrum of axion masses
and choice of initial conditions, one can readily deduce the
cosmological observables.

We found that the scalar power spectrum resulting from a generic
N-flation model is distinguishable, in principle, from the
single-field $m^2\phi^2$ inflation result. In the case of the
initial conditions that we argued are most plausible, the spectral
index can be written very simply in terms of $\beta$, as \nmpmom.
For nonzero $\beta$, the spectrum is more red than in the
associated single-field model; for the preferred value $\beta
\approx {1\over{2}}$ the change in $n_{s}-1$ is around 25\%.  Let
us stress here that there do exist parameter regimes, {\it{e.g.}}
$\beta \to 0$, in which an N-flation model is observationally
similar to $m^2 \phi^2$ inflation. However, to the extent that
these are unusual parameter choices, we find that N-flation
`usually' differs from $m^2 \phi^2$ inflation.

Because the dynamics is not independent of the initial conditions,
even at late times, we presented our general results as functions
of the initial conditions.  For example, knowing only $\beta$ and
not the initial conditions, we cannot predict the exact tilt of
the scalar power spectrum, but we can make the general prediction
that the spectrum will be more red than in the single-field case,
by an amount that depends on $\beta$.  In reasonable special cases
in which the field vevs or the energy content of each field are
roughly equal, the results are substantially simpler.  In this
connection, it would be interesting to characterize the most
plausible initial conditions for models with large numbers of
moduli.

Although we propose that random matrix theory is a very useful
tool for analyzing N-flation, we certainly do not claim that
random matrix results produce or reveal a flattening of the
inflaton potential.  These techniques merely allow us to extract a
nearly-universal axion mass spectrum in a context where direct
computation of the masses is unmanageable.  In fact, our results
imply that the scalar power spectrum of a randomly chosen
N-flation model is typically more red than in $m^2 \phi^2$
inflation. Thus, we actually find that a generic N-flation
potential is fractionally {\it{steeper}} than in the simplest
model proposed in \nf. Fortunately, this slight steepening does
not spoil slow-roll inflation, and, as we explained in \S8, it can
sometimes yield a testable prediction.

One interpretation of our results is that, in certain
circumstances, assisted inflation has a novel advantage over the
{\it{a priori}} much simpler single-field inflation models.
Suppose one chooses an embedding of inflation in string theory
that involves one or a few dynamical fields. The exact form of the
inflaton potential will, of course, depend on the specifics of the
compactification. Given an ensemble of realizations of this model,
with differing choices of the microscopic parameters, the mean
behavior will give only a qualitative guide to the behavior of a
particular realization: fluctuations from the mean are not
suppressed by large-number effects.  Consider, in contrast, an
ensemble of realizations of N-flation, or of some other sort of
assisted inflation.  The large-N simplifications of random matrix
theory, as well as the well-known suppression of fluctuations from
the mean, ensure that a typical realization of N-flation is very
well described by the mean behavior in this ensemble of
realizations.  This greatly enhances the predictivity of this
class of models.  In a sense, this runs counter to the naive
argument that the complexity of the string landscape makes it
almost impossible to draw conclusions about the inflationary
models that exist within it. One might seek other situations where
random matrix theory can  be combined with the large
dimensionality of the landscape to draw generic conclusions about
classes of models \refs{\douglas,\uday,\richardtoappear}.

In particular, it would be very interesting to find other examples
in which assisted inflation becomes somewhat more predictive and
less dependent on microscopic data as a consequence of large-N
statistical effects.  Not every assisted inflation model is
usefully specified by a matrix or a set of matrices, so random
matrix results {\it{per se}} are not always applicable.  Even so,
for a sufficiently large number of fields, statistical reasoning
may still produce useful simplifications.

We focused, for definiteness, on KKLT compactifications of type
IIB string theory, but as explained in \S9, it is entirely clear
that our conclusions generalize to other contexts.
Even so, it would be valuable to examine the detailed properties
of the N-flation potential in other string theories or in
M-theory.

We have left untouched many other interesting questions about
N-flation. In particular, although we have learned something about
the statistical properties of an ensemble of N-flation models, we
have not provided anything like a concrete realization of the
model, {\it{e.g.}} in a specific orientifold. This would be a
worthwhile endeavor.

\bigskip\centerline{\bf{Acknowledgements}}\medskip\
We thank Oliver DeWolfe, Josh Friess, Michael Haack, Simeon
Hellerman, Lev Kofman, Axel Krause, Andrew Liddle, Eugene Lim,
Juan Maldacena, Anupam Mazumdar, Hiranya Peiris, Douglas Stone,
and Jay Wacker for valuable discussions.  We are indebted to
Shamit Kachru for extremely helpful correspondence. In addition,
we thank Daniel Baumann for detailed comments on drafts of this
paper, and Shamit Kachru, Andrew Liddle, and John McGreevy for
catching an error in \S7 of a preliminary draft.  We thank the
Aspen Center for Physics, where this project was initiated, for
providing a stimulating environment. L.M. would also like to thank
the Particle Theory Group at Yale, the organizers of the Dark
Energy Workshop of the Arnold Sommerfeld Center, and the string
theory and cosmology groups at NORDITA for their hospitality. R.E.
is supported in part by the United States Department of Energy,
under contract DOE-FC02-92ER-40704. L.M. is supported in part by
the United States Department of Energy, under contract
DE-FG02-91ER-40671.


\appendix{A}{Statistical Suppression of Cross-Couplings}

In this Appendix we will compare the sizes of diagonal and
off-diagonal elements in the mass matrix \rotmassmatrix. We will
show that the diagonal elements $M_{ii}$ are normally distributed
around a mean that is larger, by a factor $\sim \sqrt{N+P}$, than
the expected magnitude $\langle |M_{ij}|\rangle$ of off-diagonal
entries $M_{ij}, i \neq j$.

As we explained in \S4, the basic structure of \rotmassmatrix\ can
be captured by \eqn\simmas{ M_{ij} = R_{iA}R^{A}_{~j} } where $R$
is a random $(N+P) \times N$ matrix that is related by a rotation
to $H_{Aj} \equiv D_{A}C_{j}$.  The entries of $R$ may have either
sign.  As we demonstrated in \S4.1, the entries of $R$ have a mean
$\mu$ that is small compared to the standard deviation $\sigma$.
In light of this property, we assume for simplicity\foot{It is
straightforward to extend our arguments, though not necessarily
the order-one coefficients, to much more general distributions.}
that the $R_{iA}$ are normally distributed around zero with
variance $\sigma^2$.

The expected value of a diagonal element $M_{ii}$ is immediately
seen to be \eqn\mij{ \langle M_{ii} \rangle = (N+P) \sigma^2 } The
off-diagonal elements are only slightly more involved. We start by
examining \eqn\odl{ M_{ij} = \sum_{A} R_{iA}R_{jA},} which
contains a sum of $N+P$ products of distinct elements
$R_{iA}\times R_{jA}$. Using the fact that the distribution of a
product of normal distributions, each with variance $\sigma^2$,
has variance $\sigma^4$, and applying the central limit theorem,
we find that the $M_{ij}$ are normally distributed with zero mean
and variance $(N+P)\sigma^4$.\foot{Because we consider a sum of
$N$ terms, not a mean, the scaling is superficially different from
the classic application of the central limit theorem.}  Thus, we
find \eqn\expma{ \langle |M_{ij}|\rangle =
\sqrt{{2\over{\pi}}}\sqrt{N+P} \sigma^2} It follows that the ratio
of the expected magnitude of the diagonal terms to that of the
off-diagonal terms is \eqn\ratio{ \langle {M_{ii}\over{|M_{ij}|}}
\rangle = \sqrt{N+P}\sqrt{{\pi\over{2}}}} which is, up to a factor
very near unity, the scaling we found in our Monte Carlo
simulation.  Thus, given the knowledge of the form of matrix
product dictated by the supergravity potential, and using our
result from \S4.1 that the $D_{A}C_{i}$ have mean much smaller
than one, we have derived the bimodal distribution found in \S5,
including the ${1\over{\sqrt{N+P}}}$ suppression of off-diagonal
elements.

\appendix{B}{Inflation in the Presence of Cross-Couplings}

In this Appendix we will study the inflationary dynamics of N
canonically-normalized fields coupled by a general quadratic
potential, in some basis in which the mass matrix is {\it{not}}
diagonal.

Let us parametrize the mass matrix as \eqn\massres{ M_{ij} =
\bar{m}^{2}\delta_{ij} + \varepsilon \bar{m}^{2} \Sigma_{ij} }
where $\varepsilon$ is a dimensionless constant, $\bar{m}$ sets
the average mass scale, and $\Sigma_{ij}$ is a symmetric matrix
whose entries have mean zero and variance $\sigma^2$.

We would like to determine the critical value of $\varepsilon$ for
which the dynamics is dominated by the diagonal terms, with no
substantial corrections due to cross-couplings.  Let us first
compute the total potential $V$ when each of the $\phi_{i}$ is
near some average value $\bar{\phi}$.  We find \eqn\visrw{ V \sim
{1\over{2}} \bar{m}^2\bar{\phi}^2\Bigl(N \pm \varepsilon N \sigma
\Bigr)} where the factor of $N$ in the final term comes from the
random walk of ${\cal{O}}(N ^2)$ entries in $\Sigma$. (By $\pm
\varepsilon N \sigma$ we are indicating the typical range of
variation of $V$.)

We now observe that \eqn\parv{ 2\partial_{i}V =  2 m_{i}^2
\phi_{i}+\varepsilon \bar{m}^2 \Sigma_{ij}\phi_{j} \sim 2 m_{i}^2
\phi_{i} \pm \varepsilon \bar{m}^2 \sigma \sqrt{N}\bar{\phi}}
where we have noted that the final term is a sum of $N$ terms
distributed around zero with variance $\sigma^2$.

With these results we can read off the slow roll parameters:
\eqn\epslam{ \epsilon_{i} = {M_{p}^2\over{2}}\Bigl( {{{2 m_{i}^2
\phi_{i} \pm \varepsilon \bar{m}^2
\sigma\sqrt{N}\bar{\phi}}}\over{{\bar{m}^2\bar{\phi}^2 N }}}
\Bigr)^{2} {\left(1 \pm {{\varepsilon \sigma}}\right)}^{-2}}

\eqn\etlam{ \eta_{ii} = {{2 m_{i}^2 M_{p}^2}
\over{\bar{m}^2\bar{\phi}^2 N }}  \left(1 \pm {{\varepsilon
\sigma}}\right)^{-1}}

 \eqn\etoflam{ \eta_{ij} = {{\varepsilon \Sigma_{ij} M_{p}^2}
\over{\bar{\phi}^2 N }} \left(1 \pm {{\varepsilon \sigma}}
\right)^{-1}}

Clearly $\varepsilon \sigma \ll 1$ suffices to arrange that
\etlam\ is unmodified by the presence of the cross-terms $\Sigma$.
Moreover, because $\langle|\Sigma_{ij}|\rangle \sim\sigma$, this
same condition on $\varepsilon \sigma$ implies that $\eta_{ij}$ is
negligible compared to $\eta_{ii}$. The most stringent constraint
is that the cross-terms do not substantially change $\epsilon$.
This requires $\varepsilon \sigma \lesssim N^{-1/2}$.

We conclude that $\varepsilon \sigma \lesssim N^{-1/2}$ is the
necessary condition for the slow-roll parameters to be unmodified
by the presence of quadratic cross-couplings $\Sigma_{ij}$.
However, we just showed above that in the case of N-flation,
$\varepsilon \sigma \sim {1\over{\sqrt{N+P}}}$.  Thus, the
cross-couplings in the quadratic-potential limit of N-flation are
naturally small enough to leave the inflationary dynamics
unaffected.  (We expect that cross-couplings cannot be neglected
in the full N-axion potential, as explained in \S3.2.)

Two comments on this issue are in order.  First, the fact that the
cross-terms are dynamically unimportant does not mean that they
should simply be omitted: the distribution of eigenvalues of the
mass matrix is different from the distribution of diagonal
elements. Second, we interpret the suppression of cross-couplings
as originating in statistical considerations, not more elementary
physical ones, as evidenced by the fact that the suppression
factor $\varepsilon$ is the square root of an integer, and not a
ratio of mass scales.

\listrefs

\end